\documentclass{aa} 

\usepackage{graphicx}
\usepackage{placeins}
\usepackage{txfonts}
\usepackage[dvipsnames]{xcolor}
\usepackage{xcolor}
\usepackage{xspace}

\usepackage[breaklinks, colorlinks, allcolors=blue]{hyperref}
\setcitestyle{citesep={,}}

\makeatletter
\renewcommand*\aa@pageof{, page \thepage{} of \pageref*{LastPage}}
\makeatother

\newcommand{\xmm}{\textit{XMM-Newton\xspace}}
\newcommand{\swift}{\textit{Swift\xspace}}
\newcommand{\chandra}{\textit{Chandra\xspace}}

\titlerunning{STONKS: Quasi-real time \textit{XMM-Newton} transient detection system}
\authorrunning{Quintin, E. et al.}

\begin{document} 

   \title{STONKS: Quasi-real time \textit{XMM-Newton} transient detection system\thanks{ The multi-mission X-ray catalog is available in electronic form at the CDS via anonymous ftp to cdsarc.u-strasbg.fr (130.79.128.5) or via http://cdsweb.u-strasbg.fr/cgi-bin/qcat?J/A+A/}} 
   \author{E. Quintin
          \inst{1},
          N.A. Webb\inst{1},
          I. Georgantopoulos\inst{2},
          M. Gupta\inst{1},
          E. Kammoun\inst{1,3,4},
          L. Michel\inst{5},
          A. Schwope\inst{6},
          H. Tranin\inst{7},
          \and
          I. Traulsen\inst{6}
          }
   \institute{IRAP, Université de Toulouse, CNRS, UPS, CNES, 9 Avenue du Colonel Roche, BP 44346, 31028 Toulouse Cedex 4, France\\
   \email{erwan.quintin@irap.omp.eu},
        \and
        Institute for Astronomy Astrophysics Space Applications and Remote Sensing (IAASARS), National Observatory of Athens, I.~Metaxa \& V.~Pavlou, Penteli, 15236, Greece
        \and
        INAF -- Osservatorio Astrofisico di Arcetri, Largo Enrico Fermi 5, I-50125 Firenze, Italy
        \and
        Dipartimento di Matematica e Fisica, Universit\`{a} Roma Tre, via della Vasca Navale 84, I-00146 Rome, Italy
        \and
        Université de Strasbourg, CNRS, Observatoire Astronomique de Strasbourg, UMR 7550, F-67000 Strasbourg, France
        \and
        Leibniz-Institut f\"ur Astrophysik, An der Sternwarte 16, 14482 Potsdam, Germany
        \and
        Institut de Ciències del Cosmos, Universitat de Barcelona, c. Mart\'i i Franquès, 1, 08028, Barcelona, Spain
}
   \date{}


  \abstract
   {Over recent decades, astronomy has entered the era of massive data and real-time surveys. This is improving the study of transient objects -- although they still contain some of the most poorly understood phenomena in astrophysics, as it is inherently more difficult to obtain data to constrain the proposed models.}
   {In order to help detect these objects in their brightest state and build synergies with multi-wavelength real-time surveys, we have built a quasi-real time automatic transient detection system for the \textit{XMM-Newton} pipeline: the Search for Transient Objects in New detections using Known Sources (STONKS) pipeline.}
   {STONKS detects long-term X-ray transient events by automatically comparing new \textit{XMM-Newton} detections to any available archival X-ray data at this position, sending out an alert if the variability between observations (defined as the ratio between the maximum flux and the minimum flux or upper limit) is over 5. This required an initial careful cross-correlation and flux calibration of various X-ray catalogs from different observatories (\textit{XMM-Newton}, \textit{Chandra}, \textit{Swift}, ROSAT, and eROSITA). A Bayesian framework was put into place to solve any ambiguous associations. We also systematically computed the \textit{XMM-Newton} upper limits at the position of any X-ray source covered by the \textit{XMM-Newton} observational footprint, even without any \textit{XMM-Newton} counterpart. The behavior of STONKS was then tested on all 483 observations performed with imaging mode in 2021.}
   {Over the 2021 testing run, STONKS provided a daily alert rate of 0.7$^{+0.7}_{-0.5}$ alerts per day, about 80\% of them corresponding to serendipitous sources. Among the detected variable serendipitous sources, there are: several highly variable active galactic nuclei (AGNs) and flaring stars, as well as new X-ray binary and ultra-luminous X-ray source candidates, some of which are present here. STONKS also detected targeted tidal disruption events, ensuring its ability to detect other serendipitous events. As a byproduct of our method, the archival multi-instrument catalog contains about one million X-ray sources, with 15\% of them involving several catalogs and 60\% of them having \textit{XMM-Newton} (pointed or slew) upper limits.}
   {STONKS demonstrates a great potential for revealing future serendipitous transient X-ray sources, providing the community with the ability to follow-up on these objects a few days after their detection with the goal of obtaining a better understanding of their nature. The underlying multi-instrument archival X-ray catalog will be made available to the community and kept up to date with future X-ray data releases.}


   \maketitle
%





\section{Introduction}


The last few decades in the field of astronomy have witnessed a marked evolution in observational methods. More and more missions have turned toward time-domain astronomy, with large frameworks aimed at performing rapid follow-ups on transient events: among them,  \textit{Zwicky} Transient Facility \citep[ZTF;][]{bellm_zwicky_2014},  SVOM mision \citep{atteia_svom_2022},  \textit{Vera C. Rubin} Observatory \citep{2019ApJ...873..111I}, and others. These missions often make use of extremely large fields of view and high return rates aimed at achieving the greatest chance for detecting a transient event. 

Because of the scarcity of X-ray photons and the need to be above the atmosphere to detect them, such all-sky monitorings have been significantly more difficult to implement in X-rays than in lower energies. Most of the current X-ray telescopes (with the exception of eROSITA and the upcoming \textit{Einstein} Probe) instead perform observations of chosen targets, placed at the center of a relatively limited field of view of a few dozen square arcminutes, with typical exposure times ranging from a few to a few hundreds of kiloseconds. Within this field of view, a number of sources will be detected that are not the target and immediate subject of the observation; these detections are referred to as "serendipitous" \citep[typically $\sim$75 per observation for \xmm, e.g.,][]{webb_xmm-newton_2020}. For most X-ray observatories, a significant effort has been put into detecting, filtering, and archiving these serendipitous sources, for which the various properties are generally summarized in the form of a catalog of detections (more details in Section \ref{sec:Catalogs}).

The available X-ray catalogs contain hundreds of thousands of detections that cover many regions of interest over several decades. Systematically exploiting them is one of the current challenges of modern X-ray astronomy. One way to make use of these catalogs is to perform a classification of the sources, either by association with other catalogs \citep[for instance][]{pineau_cross-correlation_2011} or by using more advanced probabilistic techniques \citep[for instance][]{tranin_probabilistic_2022}. Once the sources are classified, it is possible to focus on a specific type of sources and thus provide an X-ray-selected population study of these objects (e.g., \citet{vagnetti_ensemble_2011} for AGNs, \citet{2020MNRAS.491.1260S} or \citet{gurpide_long-term_2021} for ultraluminous X-ray sources, or \citet{freund_stellar_2022} for stars). 


As it gives us access to more energetic events that are often intrinsically variable, the X-ray sky is even richer in transient events than the optical sky \citep[e.g.,][]{li_populations_2022}. 
In the following paragraphs, we mention some instances of these sources and justify the interest of increasing their respective available samples. 

Tidal disruption events \citep[TDEs; e.g.,][]{gezari_tidal_2021} correspond to the disruption of a star passing within the tidal radius of a black hole due to the strength of the tidal forces; this disruption can be either complete or only partial. 
The typical expected behavior is a sudden rise in the emission of the black hole, well described by a thermal continuum, followed by a slow decay over a few years, consistent more or less with a $t^{-5/3}$ power-law decay \citep{1988Natur.333..523R}, or $t^{-9/4}$ for partial TDEs \citep[e.g.,][]{coughlin_partial_2019}. 
Surveys such as the ZTF \citep{bellm_zwicky_2014} or the All Sky Automated Survey for SuperNovae \citep[ASAS-SN;][]{kochanek_all-sky_2017} have allowed for the detection of dozens of optical TDEs \citep[e.g.,][]{hammerstein_final_2022}, while X-ray detected TDEs remain rare \citep[e.g.,][]{saxton_correction_2021}. A comprehensive list of all TDE candidates can be found in the Open TDE catalog\footnote{\url{https://tde.space/}}. A large delay between the X-ray and optical counterpart of a TDE, as seen  in ATLAS17jrp \citep{wang_discovery_2022}, could explain the observational discrepancies (as any X-ray follow-up might be too early to catch the delayed X-ray counterpart to the initial optical event). Many questions remain unanswered about the precise emission mechanisms and the multi-wavelength counterparts of these events \citep{2018MNRAS.474.3307S}. Two main points of interest about TDEs could justify the efforts of trying to find new candidates. The first advantage of TDEs is in the case of wandering IMBHs; outside of the massive flare due to the disruption of the star or a lucky lensing event, these black holes are practically undetectable. Observing TDEs in such environments is thus one of the preferred strategies for the detection of the still elusive IMBHs. The second point of interest in detecting TDEs is that the level of accretion reached during the flare goes well above the Eddington limit \citep{wu_super-eddington_2018}; the precise processes of super-Eddington accretion are still poorly understood, meaning that new samples of such processes could help us understand them.

A recently discovered phenomenon that seems to be linked to TDEs are quasi-periodic eruptions (QPEs), first discovered in 2019 \citep{miniutti_nine-hour_2019} in a past X-ray TDE \citep[GSN 069, e.g.,][]{saxton_long-term_2011,shu_long_2018}. QPEs appear as large $\sim$1h long outbursts of soft thermal X-rays, repeated every $\sim$2h--10h, with peak luminosities of $\approx 10^{42}-10^{43}\,\rm erg\,s^{-1}$ . Only six QPE sources are known to this date: \object{GSN 069}, \object{RX J1301.9+2747} \citep{sun_rx_2013, giustini_x-ray_2020}, \object{eRO-QPE1} and \object{eRO-QPE2} \citep{arcodia_x-ray_2021}, along with two additional candidates, \object{XMMSL1\,J024916.6-041244} \citep{chakraborty_possible_2021}, and Tormund \citep{quintin_tormunds_2023}. Most sources have shown a pattern in their bursts, with large and small peaks alternating; eRO-QPE1 showed a transition from such a regular pattern to a chaotic profile with overlapping peaks in less than a week \citep{arcodia_complex_2022}. The long-term evolution of GSN 069 is arguably the best contrained, with an overall decay of the emission over time, the bursts appearing only in a relatively low-flux state; a rebrightening was then observed, with the QPEs disappearing \citep{miniutti_repeating_2023}. This was followed by a new decaying phase, and the QPEs appearing again, with a different alternating pattern than before \citep{miniutti_alive_2023}. Out of the six known QPE sources, three show a link with a past TDE (GSN\,069, XMMSL1\,J024916.6-041244, and Tormund). The precise emission mechanisms at play in QPEs are still unclear. Most models invoke either specific hydrodynamical instabilities \citep[e.g.,][]{sniegowska_possible_2020, kaur_magnetically_2023, pan_disk_2022, sniegowska_modified_2023}, repeated partial tidal disruption events \citep[e.g.,][]{king_gsn_2020,zhao_quasi-periodic_2022,wang_model_2022, chen_milli-hertz_2022, king_quasi-periodic_2022}, or an inital partial TDE followed by repeated interactions between the remnant and its orbiting debris \citep[e.g.,][]{xian_x-ray_2021, linial_emri_2023,franchini_quasi-periodic_2023}. To discriminate between these models, more data are 
needed to both constrain the long-term evolution on the already-known QPE sources and to increase the sample of known QPE sources. This will allow us, for instance, to make statistically
significant population studies \citep[e.g.,][]{wevers_host_2022}. 

Another window on super-Eddington accretion is ultraluminous X-ray sources \citep[ULXs;][]{kaaret_ultraluminous_2017}. They correspond to extra-galactic, extra-nuclear sources reaching X-ray luminosities above $3\times10^{39}$\,erg\,s$^{-1}$. This somewhat arbitrary threshold was chosen as it corresponds to the isotropic Eddington luminosity of a 20 $M_\odot$ black hole \citep{remillard_x-ray_2006}. Going significantly above this value means that the source is either more massive than 20 $M_\odot$, so that  the Eddington limit can be respected. Otherwise,  it violates this limit, which means that the accretion is following a super-Eddington regime. The discovery of accelerating coherent pulsations in a ULX in M82 \citep{2014Natur.514..202B} lead to the conclusion that at least some ULXs are host to a neutron star, and thus require highly super-Eddington accretion to reach the observed luminosities (up to 500\,$L_{\rm Edd}$ for the pulsating ULX in NGC 5907 reported in \citet{israel_accreting_2017} for instance). So far, only a handful of pulsating ULXs have been found. A key feature of these known PULXs is that they seem brighter and more variable than the overall ULX population, which could hint at a physically motivated sub-classification of ULXs, or be a selection bias due to the difficulty of finding pulsations in scarce X-ray signals. Nonetheless, outstanding variability has been used as a proxy to find good candidates for pulsations \citep{2020MNRAS.491.1260S} and could allow us to detect new candidates for further pulsation search.

While the previously mentioned variable sources are extragalactic, our Galaxy is also rich in X-ray transient objects. For instance, some stars can be bright in X-rays \citep[e.g., young stellar objects,][]{preibisch_origin_2005}. Among these X-ray bright stars, some can show flaring episodes, which can be due to coronal activity for instance \citep[e.g.,][]{pallavicini_relations_1981}, or to magnetic activity \citep[e.g.,][]{stelzer_uv_2013}. These flares typically last for a few hours with peak luminosities in the $10^{29}-10^{32}$~erg~s$^{-1}$ range and are thus visible within observations of X-ray missions such as \textit{XMM-Newton} \citep[e.g.,][for a sample study]{pye_survey_2015}.

On top of TDEs, QPEs, ULXs, and stellar flares
, there is a host of other interesting X-ray variable sources: gamma ray bursts, novae \citep[e.g.,][]{konig_x-ray_2022}, cataclysmic variables \citep[e.g.,][]{webb_two_2018}, and X-ray binaries, supernovae, blazars, and changing-look active galactic nuclei \citep[e.g.,][]{graham_understanding_2020}. For all these events, an alert (and subsequent follow-up) even a week after the initial event can provide valuable information. Additionally, some newly studied variable sources are detected in other wavelengths and studying their possible X-ray counterparts might allow us to reveal or at least constrain their still unclear physical nature: fast blue optical transients \citep{margutti_embedded_2019} and fast radio bursts \citep{petroff_fast_2019}. Finally, there might even be new types of  variable unknown X-ray objects lingering in the archives that are yet to be discovered.

All of these sources are rare and show some type of variability, either in flux or spectral shape. Finding and studying them would increase their numbers and help elucidate the underlying physical mechanism governing their nature. To improve our understanding of these sources, it thus seems profitable to find new candidates, based on X-ray variability. To be able to retrieve the most constraining data for these sources, both in X-rays and in other wavelengths, it is of paramount importance to detect them when they are in their brightest state.

In this paper, we describe a new quasi-real time transient detection system that could be deployed in the \xmm~pipeline, developed as part of the XMM2Athena project \citep{webb_xmm2athena_2023}. Our approach is to compare new \textit{XMM-Newton} EPIC detections to any available archival X-ray data, in order to assess the long-term variability of the underlying object. To do this in a computationally efficient manner that would not slow down the already-existing data stream, we performed a compilation of the archival X-ray sky (through both catalogs of detections and upper-limits). This catalog-oriented approach, on top of allowing for faster computations in the pipeline, also enables various data mining endeavours in the compiled X-ray archive, the results of which have been presented in earlier publications \citep[e.g.,][]{quintin_new_2021, quintin_tormunds_2023}.

We explain the underlying multi-instrument archival catalog and archival \xmm~upper limits (Sect. \ref{sec:Method}), then describe and test the proposed transient detection system itself (Sect. \ref{sec:STONKS}), and finally discuss the main limits, expected results and future updates of this system (Sect. \ref{sec:Discuss}).

\section{Collating the archival X-ray sky}
\label{sec:Method}
\subsection{X-ray multi-instrument matching method}
\subsubsection{Data selection}

\label{sec:Catalogs}
\begin{table*}[]
\resizebox{\textwidth}{!}{
\begin{tabular}{c|c|c|c|c|c|c|c|c}
Telescope & Catalog & Sky coverage & Limiting sensitivity & Spatial resolution  & Sources & Detections & Dates & Reference\\ 
           & & (sq. degrees) & (erg~s$^{-1}$~cm$^{-2}$)   &  (FWHM arcsecond)  & & & \\\hline \hline
   \rule{0pt}{1.\normalbaselineskip}\textit{XMM-Newton} & 4XMMDR11 & 560 & $\sim10^{-15}$ & 5 & 470 000 & 700 000 & 2000--2020 & \citet{webb_xmm-newton_2020}\\
           &  4XMMDR11s  & 560 & $\sim10^{-15}$ & 5 & 34 000+ & 51 000+ & 2000--2020& \citet{traulsen_xmm-newton_2019}\\ 
           &  XMMSL2 & 65 000 & $\sim10^{-12}$ & 10 & 22 000 & 27 000 & 2001--2014 & \citet{saxton_first_2008}\\ \hline   
   \rule{0pt}{1.\normalbaselineskip}\textit{Swift}   &  2SXPS & 3 790 & $\sim10^{-13}$ & 6 & 145 000 & 300 000 & 2005--2018& \citet{2020ApJS..247...54E}\\\hline
   \rule{0pt}{1.\normalbaselineskip}\textit{Chandra}   &  CSC 2.0 & 550 & $\sim10^{-16}$ & 0.75 -- 5 & 200 000 & 300 000 & 2000--2014 & \citet{evans_chandra_2020}\\\hline
   \rule{0pt}{1.\normalbaselineskip}ROSAT & RASS & 41 000 & $\sim10^{-12}$ & 20 & 60 000 & 60 000 & 1990--1991 & \citet{boller_second_2016}\\
      &  WGACAT & 7 500 & $\sim10^{-13}$ & 20 & 70 000 & 80 000 & 1991--1994 & \citet{white_wgacat_1994}\\\hline
   \rule{0pt}{1.\normalbaselineskip}eROSITA &  eFEDS & 140 & $\sim10^{-14}$ & 5 & 20 000 & 20 000 & Nov. 2019  & \citet{salvato_erosita_2022}
\end{tabular}
}
\caption{Properties of the catalogs after quality filtering. The limiting sensitivities are typical flux values in the corresponding instrument's energy band (see Fig.\,\ref{fig:EnergyBands}), but numerous instrumental effects (off-axis angle, background, exposure time) will impact this value. For \textit{Chandra}, the two values for spatial resolution correspond to the on-axis and 10' off-axis FWHM. For the \textit{XMM-Newton} Stacked catalog, we only show the number of new sources and number of new detections (which might be associated with already known sources). }
\label{tab:CatProperties}
\end{table*}

\begin{figure}
    \begin{center}
    \includegraphics[width=\columnwidth]{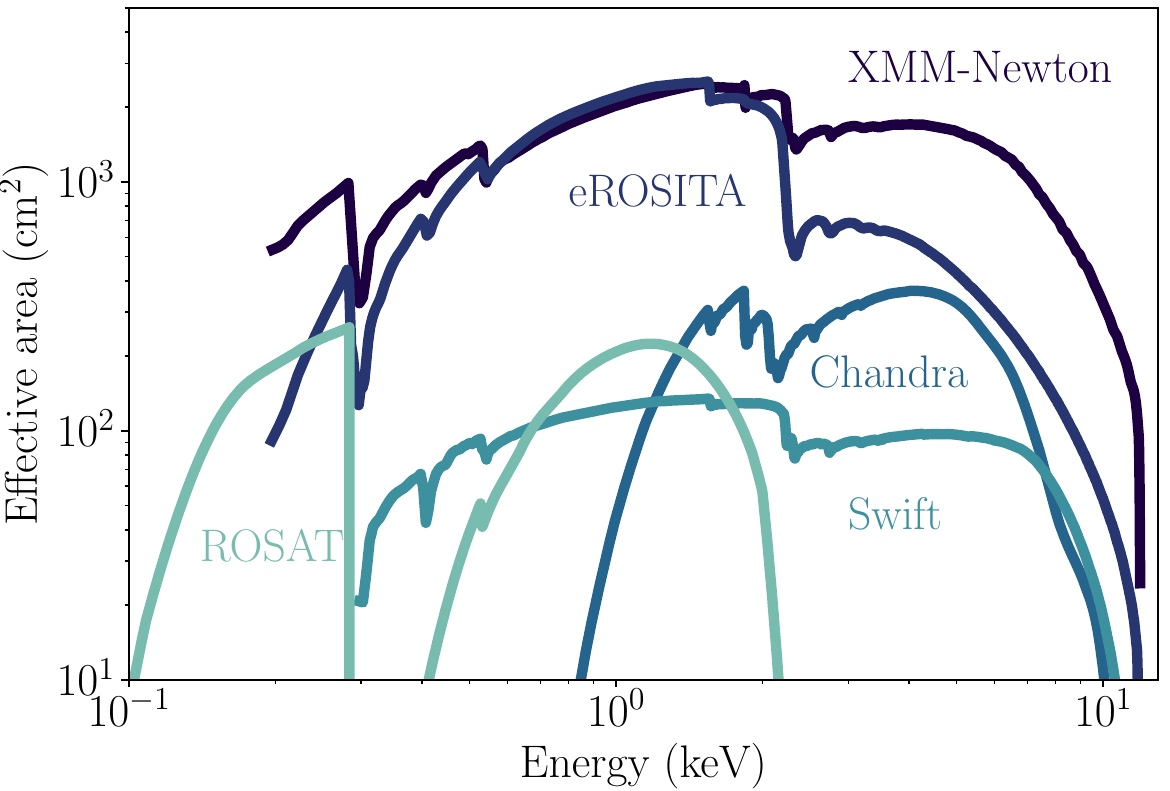}
    \caption{Comparison of effective areas of all the X-ray missions used in this work. For \xmm~we show the combined EPIC effective area. For \chandra,~we show the ACIS-I effective area as of 2022. For \swift~we show the XRT effective area. For eROSITA we show the effective area from the combined seven telescopes. For ROSAT we show the PSPC effective area.}
    \label{fig:Areas}
    \end{center}
\end{figure}

Some studies have been performed to systematically look for variable objects in the archive of some X-ray observatories \citep[e.g., the search for fast X-ray transients in the \textit{Chandra} archive or the EXTraS project for \textit{XMM-Newton};][]{jonker_discovery_2013,luca_extras_2021}. However, in order to improve our chances of finding long-term variability in serendipitous sources, a multi-instrument approach is preferable, as it provides an increased number of data points for a given source. For this reason, we used eight different X-ray catalogs, with complementary strengths and weaknesses. This method is similar for instance to the HILIGT web service \citep{saxton_hiligt_2022,konig_hiligt_2022}. A summary of the catalogs' respective properties can be found in Table\,\ref{tab:CatProperties} and their effective areas are shown in Fig.\,\ref{fig:Areas}.

The first three catalogs we chose are 4XMM DR11 \citep{webb_xmm-newton_2020}, 2SXPS \citep{2020ApJS..247...54E}, and 2CXO \citep{evans_chandra_2020}, which are the source catalog respectively for \xmm, \textit{Swift}/XRT, and \chandra. Their respective sensitivity, angular resolution and sky coverage (see Table\,\ref{tab:CatProperties}) differ significantly because of the different technical setups of their instrumentation, driven by different scientific goals.


We also took into account two additional catalogs obtained from XMM-Newton: the slew catalog XMMSL2 \citep{saxton_first_2008} and the stacked catalog 4XMM DR11 Stacked \citep{traulsen_xmm-newton_2019}. The first one corresponds to detections obtained during the slewing of the instrument, between two consecutive pointings. It provides us with a large sky coverage, at low exposure times and thus low sensitivity. The second catalog is obtained from the stacking of overlapping observations, which provides improved sensitivity and more reliable source parameters compared to single observations, as well as possibly new detections in some observations. For the stacked catalog, we only kept detections that were not in the initial pointed catalog (corresponding either to sources that are in the initial catalog but for which some observations did not lead to clean detections, and also for entirely new sources absent from the initial catalog).

We added two ROSAT catalogs, 2RXS \citep{boller_second_2016} and WGACAT \citep{white_wgacat_1994}, corresponding respectively to the sky survey and to subsequent pointed observations. Despite their relatively low sensitivity and angular resolution, these catalogs are very useful for their wide sky coverage, as well as for the fact that they provide us with a longer temporal baseline to study variability.

Finally, the study of long-term variability of X-ray sources will be immensely improved by the data from eROSITA \citep{predehl_erosita_2021}, which will provide multiple all-sky X-ray surveys with sensitivity levels comparable to that of \xmm. In order to make a proof of concept of the interest of using future eROSITA data within our framework, we have used the available early data from the eROSITA Final Equatorial Depth Survey catalog \citep[eFEDS;][]{salvato_erosita_2022}, which covers a small patch of the sky of about 140 square degrees, with non-contemporaneous \xmm~and \chandra~observations. \citet{boller_erosita_2022} have already performed a study of the variable sources in eFEDS, although our method should reveal additional long-term variability.

Once selected, these catalogs have been cleaned using different selection criteria with the aim of keeping only point-like sources, avoiding spurious detections and improving the overall quality of the final catalog. The cleaning procedures were performed on detections; the remaining sources are those that have at least one remaining clean detection. The various catalog-specific selection criteria are summarized in Appendix\,\ref{sec:AppSelectionCriteria}.

\begin{figure*}
    \begin{center}
    \includegraphics[width=\textwidth]{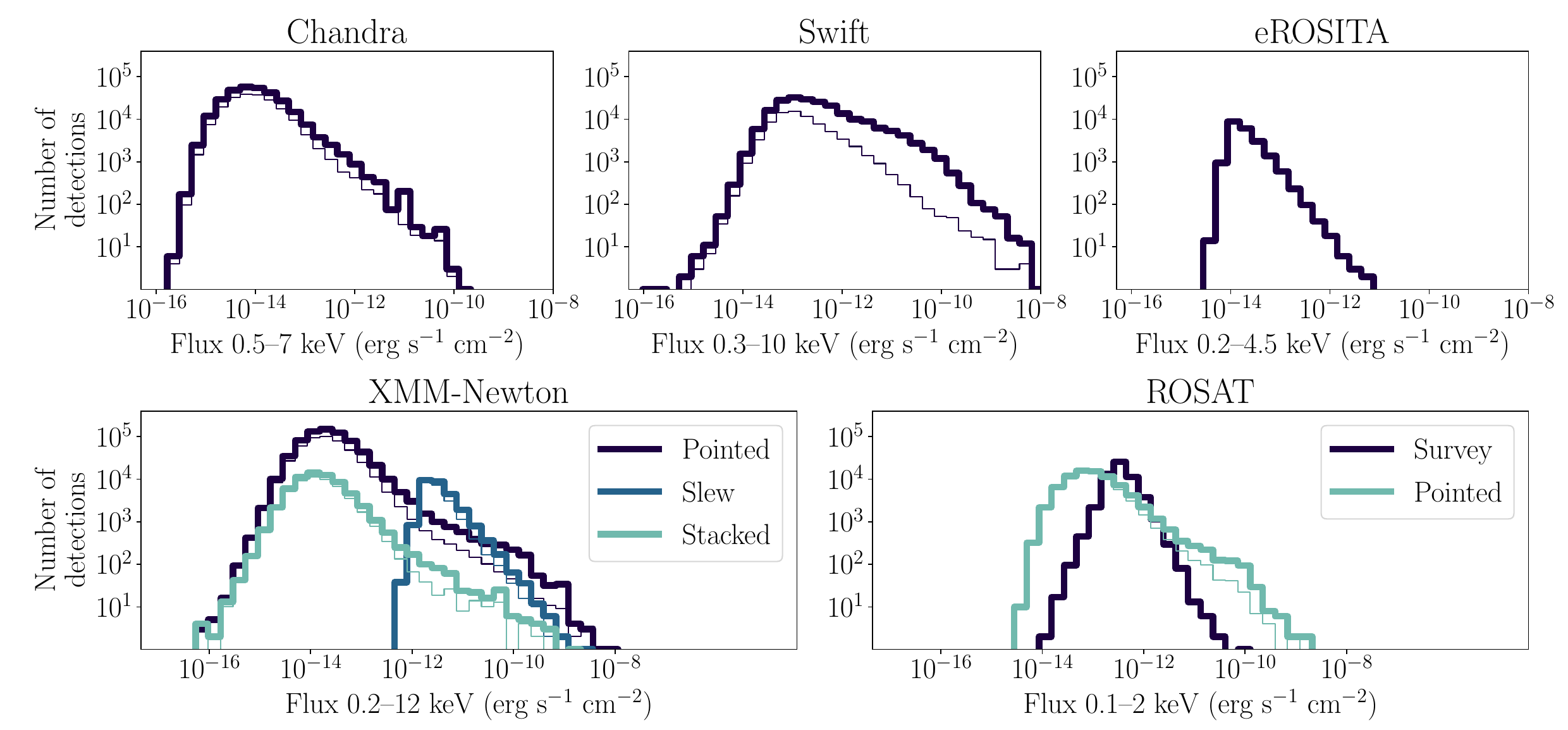}
    \caption{Flux distributions of each X-ray observatory used in this study, in their native energy band, with the different catalogs shown in different colors. For each catalog, we show the flux distribution of all detections (thick line), as well as the flux distribution averaged for each source (thin line). The difference between the detection-wise and source-wise flux distributions depends on the observational strategy of each X-ray instrument.}
    \label{fig:Flux}
    \end{center}
\end{figure*}

The resulting flux distributions of each catalog are shown in Fig.\,\ref{fig:Flux}. In particular, this figure shows the flux distribution of all detections, as well as the flux distribution averaged for each source. The shape of these distributions and the differences between them will depend on the overall observing strategy -- for instance, the \swift~flux distribution loses a significant fraction of its high-flux component when averaging over each source, because \swift~is often used as a monitoring telescope for bright objects.

Once these quality checks have been applied, we have a total of about 1 million X-ray catalog sources and 1.5 million detections. For each detection, we have a rate in the corresponding total energy band of the instrument, as well as in different sub-bands that will be used to access spectral information. We now need to associate those sources together. This will be done by matching the catalogs two by two at first, in order to take into account their respective astrometric differences and avoid the combinatorial difficulties of a single multi-catalog match; then, these two-by-two matches will be combined into multi-catalog sources, using a conservative fusion approach.

\subsubsection{Two-by-two catalog matches}
The core of our method is based on the two-by-two correlations between catalogs. These were performed using \texttt{STILTS} \citep{taylor_stilts_2006}, based on the positions and $3\sigma$ circular position errors for each source in the two considered catalogs. Among all combinations of catalogs, we did not compute the \xmm~pointed to \xmm~stacked cross-correlation, as this work was already performed and manually screened in the elaboration of the \xmm~stacked catalog \citep{traulsen_xmm-newton_2019}. Two issues arose from this naive cross-matching method.

The first issue we encountered was for very bright X-ray sources ($F\sim10^{-10}$ erg\,s$^{-1}$). For these sources, the large number of photons allowed for a very precise fit of the PSF; so precise in fact that the $3\sigma$ positional errors can be smaller than the astrometric error between catalogs, thus preventing the matches for bright sources. To prevent this, we have computed an estimation of the astrometric error for each catalog combination, by producing a naive correlation and taking the closest match for each source using a very large position cutoff (1 arcmin). Assuming that the coordinate differences follow the same normal distribution, the angular distance distribution of this naive match should yield a Rayleigh distribution at close distance, with an excess at large distance due to spurious associations \citep[this method was used for instance in][]{boller_second_2016}. Taking the maximum of this Rayleigh distribution allows us to retrieve its $\sigma$ value, which roughly corresponds to the standard deviation of the coordinate errors. For the ulterior matches between those two given catalogs, the matching distance was taken as the maximum between the $3\sigma$ position error and the estimated astrometric error.

The second issue arises for ambiguous correlations. Indeed, taking the $3\sigma$ positional error and the astrometric error into account can lead to a reasonably large maximum matching distance, that can then lead to a number of possible counterparts. In this case, the \texttt{STILTS} command will return a group of ambiguous associations, with all allowed combinations of source associations. Identifying the correct counterpart for each source is essential, as spurious associations may lead to large, erroneous variability. For this purpose, we have developed a Bayesian approach to quantify the quality of an association, which will allow us to compare between candidates and decide whether the match is decisive or unclear. The precise method is similar to the one implemented in \texttt{NWAY} \citep{2018MNRAS.473.4937S}, which was inspired from \citet{2008ApJ...679..301B}. We denote $H_i$ as the hypothesis that the $i^{th}$ possible match between two catalog sources is real, and $\bar{H_i}$ as the opposite hypothesis; the data, namely, the position and position error of each source, are noted as $D_i$. The Bayesian probability for the $i^{th}$ match is thus:
\begin{equation}
    P(H_i|D_i) = P(D_i|H_i)\times \frac{P(H_i)}{P(D_i)}
.\end{equation}
The end goal will be to compute the ratio of this value between different counterparts, $i$. With a flat prior on the data and $P(H_i)$ only depending on the overlap between two catalogs and thus independent of $i$, for a given catalog combination the only value of interest is $P(D_i|H_i)$. With the same assumptions as the Appendix B from \citet{2008ApJ...679..301B} (i.e., a spherical normal error on position, with error bars and distances small compared to the size of the sky), this value is given by: 
\begin{equation}
    \label{eq:B}
    P(D_i|H_i) = \frac{2}{\sigma_1^2 + \sigma_2^2} \text{exp}\Bigg(-\frac{\psi^2}{2(\sigma_1^2 + \sigma_2^2)}\Bigg)
,\end{equation}
with $\sigma_1$ and $\sigma_2$ the error bars of the two associated sources and $\psi$ the angular distance between their positions; at this stage, the astrometric error is not taken into account. We compute this "association score" for all associations, and use it as a way to compare between ambiguous ones. After manual screening, we take a ratio of 3 between two scores as a very good indication that one association is favored over the other; a ratio below that generally corresponds to different spatial resolutions resulting in two sources for an instrument being seen as a single source for another instrument (\textit{Chandra} vs. \textit{Swift} typically).

The precise workflow for each two-by-two catalog correlation is thus as follows: we first estimate the astrometry error between two catalogs by performing a crude correlation, and taking its typical angular distance; we perform the precise correlation using 3$\sigma$ positional errors and astrometric error; the association score for all associations is computed following Eq. \ref{eq:B}. Then, for each group of ambiguous associations, we sort by order of association score. We compare the score of the most probable association of the group to the score of the second most probable association involving any of the two concerned sources (this is the major difference with \texttt{NWAY}, in which only the possible matches for one source of the pair are considered). If the ratio is higher than 3, we validate the first association and ignore all the other ones; else, we ignore all the associations for these two sources, as it is impossible to safely conclude on the association. Finally, we proceed until all combinations have been either accepted or ignored.

Deviating from \citet{2008ApJ...679..301B}, we do not include photometric information in our Bayesian approach, because a photometry-based match relies on constant flux assumption, while we search for transients. One issue that may arise from this choice is to favor a close spatial match between a bright and a faint source from two catalogs, where one of them has poorer spatial localisation (e.g., ROSAT or \textit{XMM-Newton} slew), while the correct bright (non-variable) match is not favored spatially. This can be avoided by using the ambiguous match solver, which will be able to flag such situations. This can also be manually treated at the quality check step (see Sect. \ref{sec:STONKS}).

\subsubsection{Combined catalog matches}
Once all two-by-two correlations of catalogs are performed, we need to merge these into multi-catalog associations. This requires dealing with associations that are inconsistent between catalogs. We chose a conservative approach, in which chain-like correlations are refused (i.e., with three sources from catalogs A, B, and C, source B is associated with both A and C, but A and C are only associated with B and not with each other). To do this, we first classify the catalogs in an arbitrary order of interest, with the idea that such chains will be dealt with in order of priority (i.e., sources A and B first in the previous example). In a pair of catalogs, the first is hereafter called primary, the other secondary. We compute all two-by-two correlations for the primary catalog with any secondary catalog, including solving ambiguous correlations using the association score, as presented in the previous section. For each source from the primary catalog, we validate its associations with all its corresponding secondary sources into the final multi-instrument catalog. At this stage, we should have recovered any counterpart to each source of the primary catalog. We then reiterate this procedure by promoting the secondary catalog to primary. However, an additional condition to accept an association now is that neither the (new) primary, nor the secondary sources, have already been encountered at a previous stage in this procedure. If they had already been encountered, this means that they are either already part of a validated association, or part of a chain-like association, which is prohibited. We proceed with this, until all two-by-two catalog correlations are merged into a single multi-catalogs catalog, where associations are performed both conservatively and quantitatively, through the use of the Bayesian association score.

\subsection{Cross calibration}
\label{subsec:FluxCal}
\begin{figure*}
    \centering
    \includegraphics[width=\textwidth]{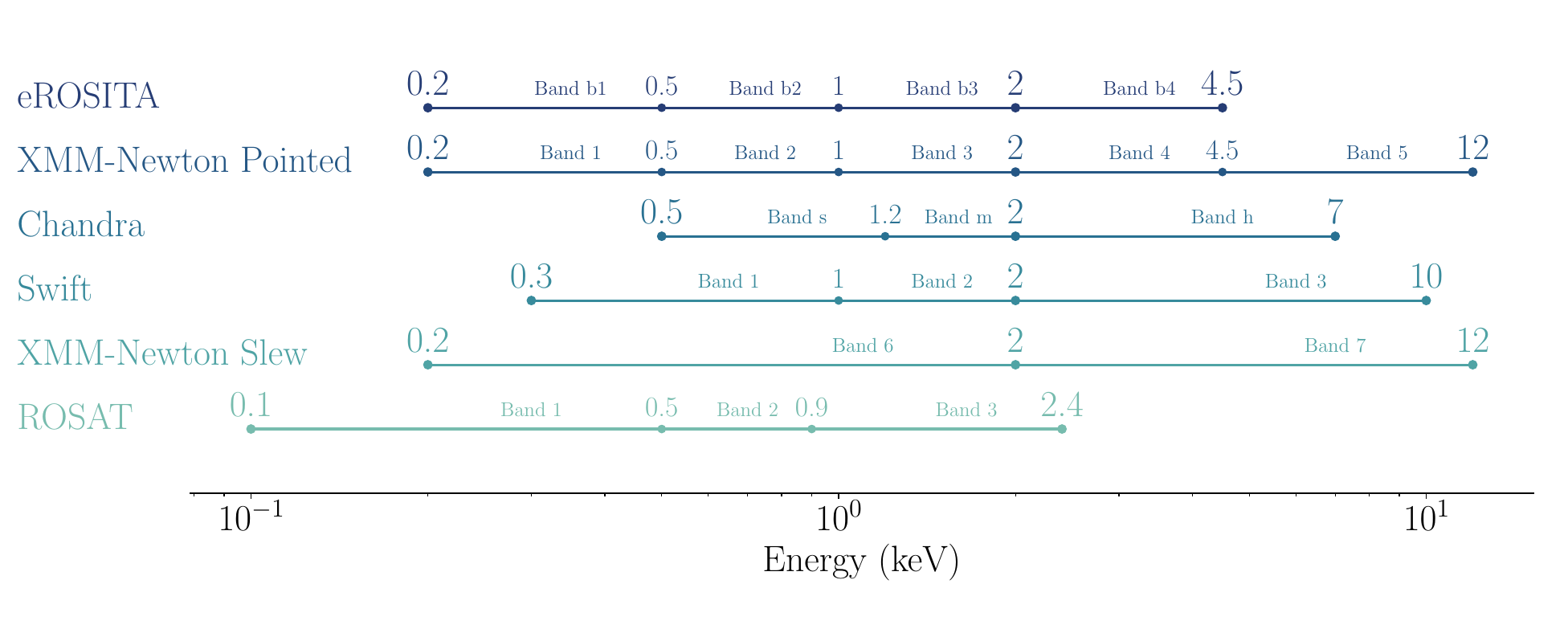}
    \caption{Energy bands of the various catalogs and instruments used in this work. We also show the catalog-specific internal energy bands, with their catalog name indicated above their respective energy regime.}
    \label{fig:EnergyBands}
\end{figure*}

Once sources are associated in the multi-instrument catalog, we need to compare the various fluxes of each catalog source. However, reliable cross-calibration of the various instruments is a major challenge for any multi-catalog flux comparison. Each instrument has a different response (see Fig.\,\ref{fig:Areas}). While most of those instrumental effects are taken into account by the processing pipelines through ancillary and response files, some biases remain \citep[of the order $\sim$8\% between the EPIC instruments for instance,][]{smith_mjs_xmm-soc-cal-tn-0018_2022}, and about 5-15\% between different missions when working in the same energy band \citep[e.g.,][]{madsen_iachec_2017}. However, the energy bands differ between the missions. Figure \ref{fig:EnergyBands} shows the respective total energy bands of each specific catalog, as well as the catalog-dependent internal energy bands. A useful feature one can see in this figure is that, for all catalogs, the value of 2 keV is a limit to some internal bands. 

To compare the fluxes obtained by different instruments and assess the source's variability, we first need to convert each detection to a single, common energy band; we cannot directly compare for instance the \xmm~flux of a source in the 0.2-12 keV band, to that of \textit{Chandra}, which is optimised in the 0.5-7 keV band. The common band we chose to compute fluxes is the 0.1-12 keV band, as it allows us to constrain the energy bands of every one of the missions we used (\xmm~ going to the highest energies and ROSAT to the lowest). Then, to extrapolate the instrument detections to this common band, we need to assume a specific spectral shape. We chose an absorbed power-law, of parameters $\Gamma=1.7$ and $N_{\rm H}=3\times10^{20}$cm$^{-2}$. The reason this was chosen is that these parameters correspond to a typical X-ray source \citep[e.g.,][]{watson_xmm-newton_2009}, and the resulting spectrum is thus rarely far from the actual spectrum of the source -- for this reason, it was used to compute fluxes for instance in the \xmm~and \swift~catalogs. Any other spectral model would not be self-consistent with the direct use of the catalog fluxes (which use this assumption), and would thus require further calibration. Assuming this fixed spectral shape, the contributions to the total flux of each band as well as the fraction of the flux missed by each instrument is shown in Table \ref{tab:EnergyConvFactor}.

\begin{figure}
    \centering
    \includegraphics[width=\columnwidth]{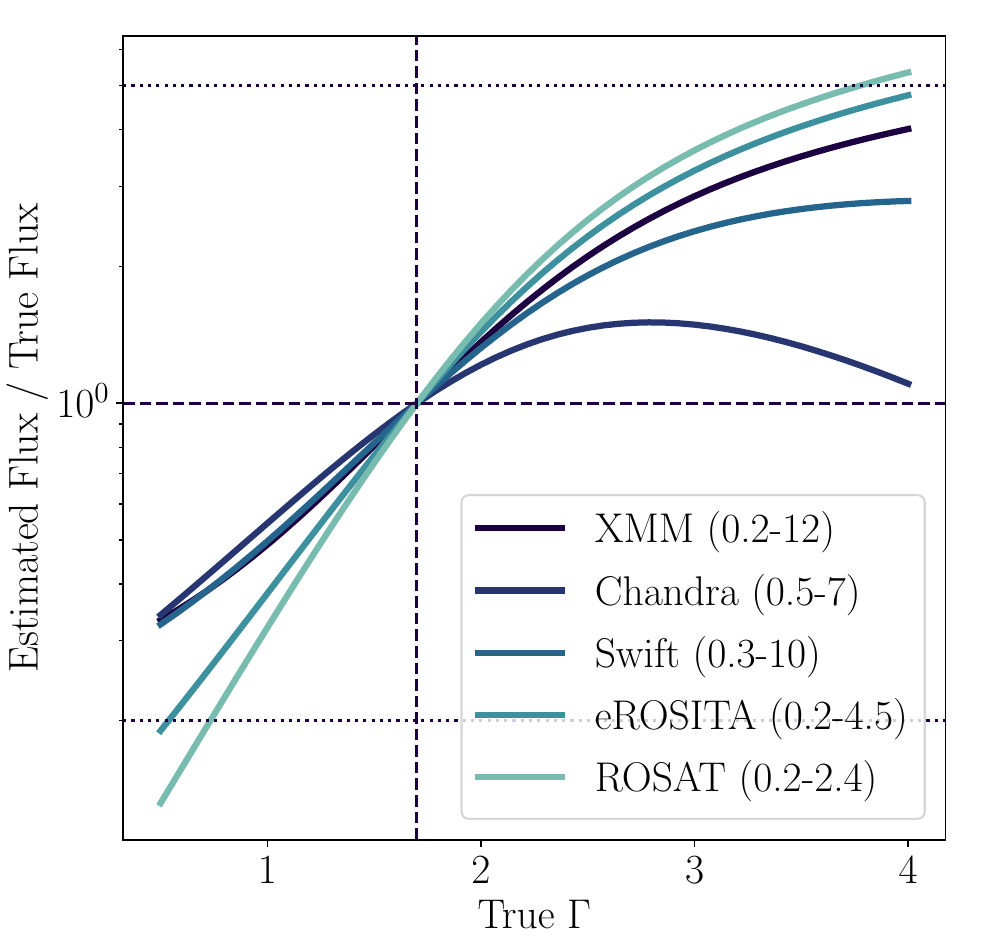}
    \caption{Evolution of the ratio between the flux extrapolated from each mission band assuming $\Gamma=1.7$ and $n_{H}=3\times10^{20}$~cm$^{-2}$, and the true flux of a source, depending on the value of its photon index $\Gamma$. The dashed lines correspond to the reference ($\Gamma=1.7$ and ratio of 1), and the dotted lines correspond to a factor of 5. While ROSAT goes over the threshold of 5 for the softest sources, what matters most to our study is that at a given $\Gamma$ the ratio between different missions is below five (to avoid spurious alerts).}
    \label{fig:SpectralAssumption2}
\end{figure}

\begin{figure*}
    \centering
    \includegraphics[width=\textwidth]{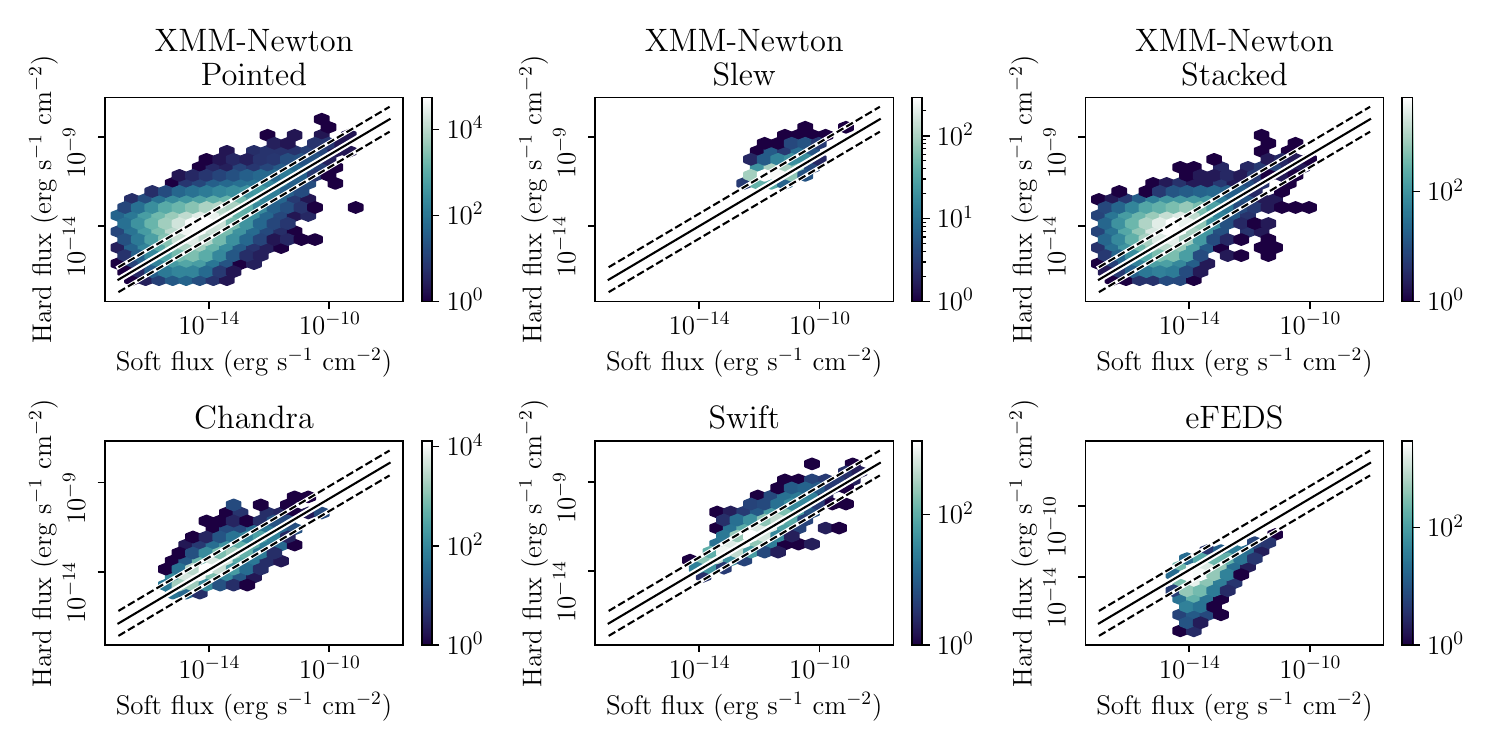}
    \caption{Comparison between the hard and soft fluxes for each mission with hard detections (i.e., >2 keV). The black lines show the expected behavior of the spectral assumption (absorbed power law of $N_{\rm H}=3\times10^{20}$cm$^{-2}$ and $\Gamma=1.7$), and the black dotted lines show a departure by a factor of 5 from this behavior. While the spread around the assumed shape can appear significant, it is important to remember that the error bars on these hard and soft fluxes are significant as well (typically signal to noise ratio of about 3 or less), so the statistical significance of the spread is reduced}
    \label{fig:HardVsSoft}
\end{figure*}

This spectral shape assumption has its limits. It fits relatively well to the majority of sources, however, for the softest or hardest sources there can be some discrepancy. Figure \ref{fig:HardVsSoft} gives the distribution of the soft vs. hard fluxes (<2keV vs >2keV) for each detection in the instruments with a hard energy band (i.e., not ROSAT
). Any departure from the black line means a departure from the assumed spectral model. To validate the use of this spectral assumption in order to assess variability between detections of different instruments, it is necessary to estimate the spurious variability that would appear from wrongfully extrapolating the source's flux beyond the specific instrumental bands. For this purpose, we implement two tests. The first test of validity of our spectral assumption simply consists in computing the error in flux estimation arising from this assumption, depending on the source's true spectral shape. In practice, we compute the evolution of the extrapolated flux from each mission's band to the total band assuming a fixed $\Gamma=1.7$ and $n_{H}=3\times10^{20}$~cm$^{-2}$, depending on the actual photon index of the source (in a 0.5--4 range). A photon index of $\sim$4 is reasonably close to a soft thermal emission, at least from a catalog point of view. The various fluxes were computed using JAXspec \citep[][Dupourqué et al. in prep.]{barret_simulation-based_2024}. The results can be seen in Fig. \ref{fig:SpectralAssumption2}. In this figure, one can see that in this range of photon indices, while the spectral assumption indeed leads to a bias on the estimated flux, this bias stays overall below a factor of five. More importantly, the respective biases of different missions stay closer than a factor of five from each other, which means that at a given value of $\Gamma$, the calibration method should lead to a minimal number of spurious alerts.
To assess the effect of such extrapolation on data rather than theoretical spectra, we test it on the \xmm~data, and analyse the variability that is created solely from this method. We started by truncating the energy bands of \xmm~to fit those of \textit{Chandra}, which is the second most delicate extrapolation after ROSAT. For each \xmm~detection, we removed the first and last bands, to retrieve \xmm~fluxes in the 0.5--4.5 keV. To get the same higher energy limit, namely, 7 keV for \chandra, we had to extrapolate the flux of the \xmm~band 4 from 2--4.5 keV to 2--7 keV. This extrapolation is done using the spectral assumption of an absorbed powerlaw, with the aforementioned parameters. The effect of this assumption on a single band is much smaller than on the entire \xmm~energy bandwidth, and is thus neglected. After this, we extrapolate the simulated 0.5--7~keV flux to the 0.1--12~keV band using the same conversion factor we would have used for \chandra~data. Comparing the resulting flux to the actual 0.2--12 keV \xmm~detection allows us to assess the spurious variability caused by this spectral approximation (the 0.1--0.2 keV contribution is negligible in this spectral assumption). We use a conservative estimate of the variability between the two flux computations. We compute the ratio of the higher of them minus its error over the lower flux plus its error:
\begin{equation}
    V_{\rm Conservative} = \left\{
    \begin{array}{ll}
        max\left(\frac{F_{\rm Band~8}-\sigma_{\rm Band~8}}{F_{\rm Extrap.}+\sigma_{\rm Extrap.}},1\right) & \mbox{\rm if~} F_{\rm Band~8}>F_{\rm Extrap,}\\
        min\left(\frac{F_{\rm Band~8}+\sigma_{\rm Band~8}}{F_{\rm Extrap.}-\sigma_{\rm Extrap.}},1\right) & \mbox{\rm if~} F_{\rm Band~8}<F_{\rm Extrap.} 
    \end{array}
    \right.
\end{equation}
with $F$ and $\sigma$ the respective flux and $1\sigma$ flux errors for both methods. This estimate takes the value of 1 in the case both methods are consistent at a $1\sigma$ level, and otherwise takes the most pessimistic assumption for variability. This metric was used because it is similar to the one used later on for variability alerts (see Eq. \ref{eq:variabilityDef}), a source being labeled as variable if this metric is above 5 (or here below 0.2 as well).

\begin{figure*}
    \centering
    \includegraphics[width=\textwidth]{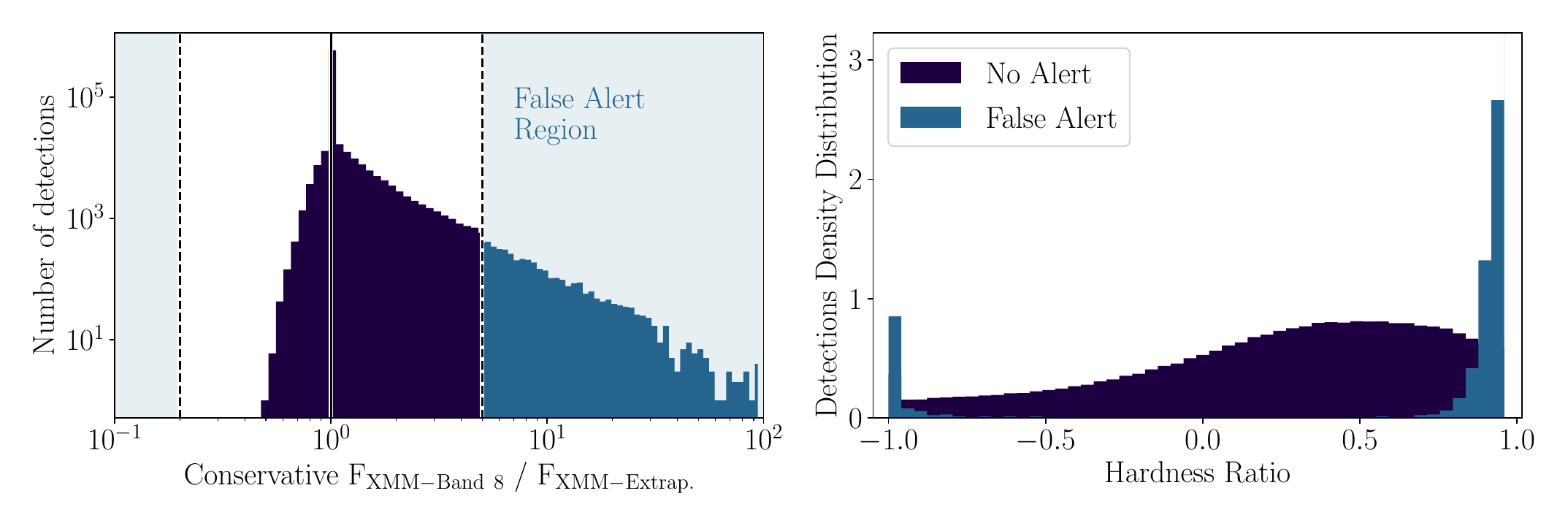}
    \caption{Assessment of the effect of the spectral assumption on variability estimates. \textit{Left Panel:} Distribution of the conservative estimate of the variability between the true flux, and the one obtained after cropping to the \chandra~bandwidth and extrapolation to the 0.1--12 keV band. All detections with a variability larger than a factor of 5 between both methods would lead to spurious transient alerts. \textit{Right Panel:} Comparison between the hardness ratio density distributions of the detections that lead to spurious alerts (light blue) and the ones without alerts (dark blue). This confirms that spurious alerts can happen in the case where the spectral assumption does not fit the data well, that is, for extreme hardness ratios.}
    \label{fig:SpectralAssumption}
\end{figure*}

The resulting spurious variabilities can be seen in  Fig.\,\ref{fig:SpectralAssumption}. We retrieved about 4 000 spurious alerts out of the 700 000 detections, amounting to about 0.6\% false alert rate. These alerts are indeed caused by the softest and hardest sources of the catalog, for which the assumption does not hold well -- this can be verified in the right panel of Fig.\,\ref{fig:SpectralAssumption}, showing the difference in density distribution of hardness ratios of the false alert detections. 

This spurious alert rate is reasonably small, however the total alert rate being about $\sim$2.5\% of detections (see Sect. \ref{sec:Testing}), this leads to a contamination of the alerts by at most $\sim$20\% and could warrant further attention. While a more adaptive spectral approximation would be possible (e.g., based on the measured hardness ratio), this solution would be very biased for low signal-to-noise detections, that tend to be significantly harder or softer than bright detections purely because of statistical effects. This would in turn dramatically increase the false alarm rate for faint detections, which is not desirable. Additionally, a minority of detections from the multi-instrument archives have available hardness information (e.g., only $\sim$20\% of both the \textit{Chandra} and \textit{Swift} archives). Overall, proper spectral data is simply not widely available in a purely catalogue-oriented approach, and a common spectral assumption is justified (which is why this solution is already implemented for each respective catalog). Alternative methods for flux extrapolations, using additional data not present in the catalogs, will be explored in the future \citep[e.g., using the archive-wide spectral fitting computed for \textit{XMM-Newton} as part of XMM2Athena,][]{webb_xmm2athena_2023}.  For now, we put in place different safeguards to warn and help the user in the case of a possible failure of this assumption, presented in Sect. \ref{sec:STONKS}.

\subsection{Upper limits}

Correlating the sources from several catalogs allows us to retrieve the flux evolution of a given physical source between several epochs. The main use case of this method is when the source was detected in the different catalogs individually. However, this method also allows us to uncover valuable information in the case where it was observed but not detected by one of the instruments. Indeed, the fact that a source was within the field of view of a given observation but not detected means that it was, at the moment of the observation, below the sensitivity of the used detection method at this point of the instrument. By computing the said sensitivity, we can retrieve an upper limit on the source's flux. This phenomenon takes place in two instances: either its intrinsic flux is constant and the observation in which it was detected previously has a better sensitivity than the one that missed it; or, the source is transient. 

We put this idea into practice for the \xmm~upper limits. We selected two types of sources for the upper limits computation: the first type of sources are known, detected-at-least-once \xmm~sources. This allows us to check whether these known \xmm~sources were detected every time they were observed, which is a piece of information absent from the \xmm~base catalog, but present in the \xmm~stacked catalog. 
The second type of source for which the \xmm~upper limits are relevant are for the sources only present in other catalogs, but that have been observed by \xmm. Using the 4XMM DR11 Multi-Order-Coverage map \citep[MOC;][]{fernique_moc_2014} which provides us with the spatial footprint of the observations, we selected all mutli-catalog sources that lie within this MOC but are far away ($>10"$) from any \xmm~source. This was done using the MOCPy package \citep{boch_mocpy_2019}. For all those sources, the upper limits were computed using RapidXMM \citep{ruiz_rapidxmm_2022}. We only kept the upper limits with a 0.2--12 keV quality flag of 0, and that were not simultaneous with a \xmm~stacked detection. We then converted the obtained 1$\sigma$ count-rates upper limits to 0.2--12 keV flux upper limits, using the same spectral assumption of a power-law of photo-index $\Gamma=1.7$ and $N_{\rm H}=3\times10^{20}$ cm$^{-2}$. While the RapidXMM framework provides pre-computed upper limits for all three EPIC instruments individually, we used the mathematical framework presented in \citet{ruiz_rapidxmm_2022} to compute the EPIC combined $3\sigma$ flux upper limits, in order to obtain more constraining upper limits. 

Additionally, we used upper limit information from both \textit{Chandra} and \textit{Swift}, but only for their respective sources. For \textit{Chandra}, the non-detections of \textit{Chandra} sources are directly available in the catalog. For \textit{Swift}, the upper limits are not readily available in a catalog-based approach, but we have access to the stacked detections. They correspond to the average flux for a source over all its \textit{Swift} exposures, and also provide us with the dates for the first and last observations. Thus, any \textit{Swift} detection that is significantly above a stacked \textit{Swift} flux hints at variability (for an example, see Fig. \ref{fig:MisclassifiedULX} or Fig. \ref{fig:NewXRB}).

\subsection{X-ray multi-instrument catalog properties}
\subsubsection{Matching statistics}

\begin{figure}
    \centering
    \includegraphics[width=\columnwidth]{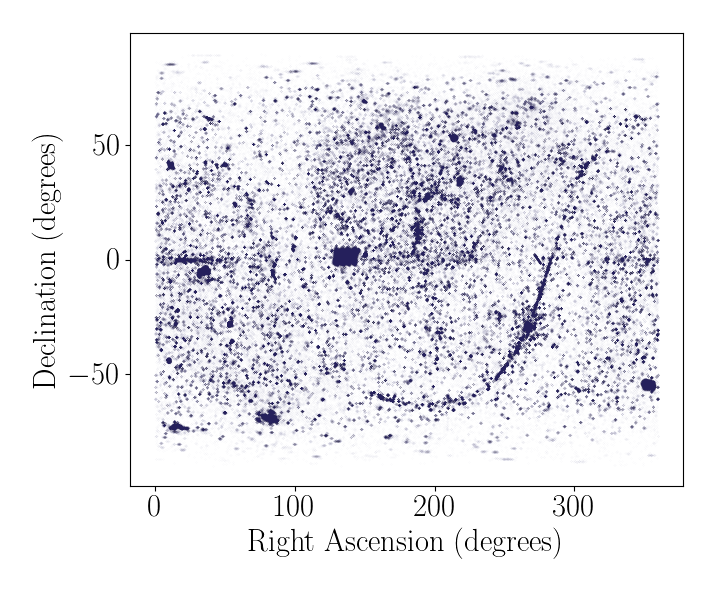}
    \caption{Sky map of the multi-instrument catalog. The galactic plane is visible, as well as the eFEDS field of view around R.A $\sim$130$^{\circ}$ and Dec. $\sim$0$^{\circ}$. This shows the inhomogeneity of the archival X-ray sky coverage.}
    \label{fig:Coverage}
\end{figure}

\begin{figure}[h]
    \centering
    \includegraphics[width=\columnwidth]{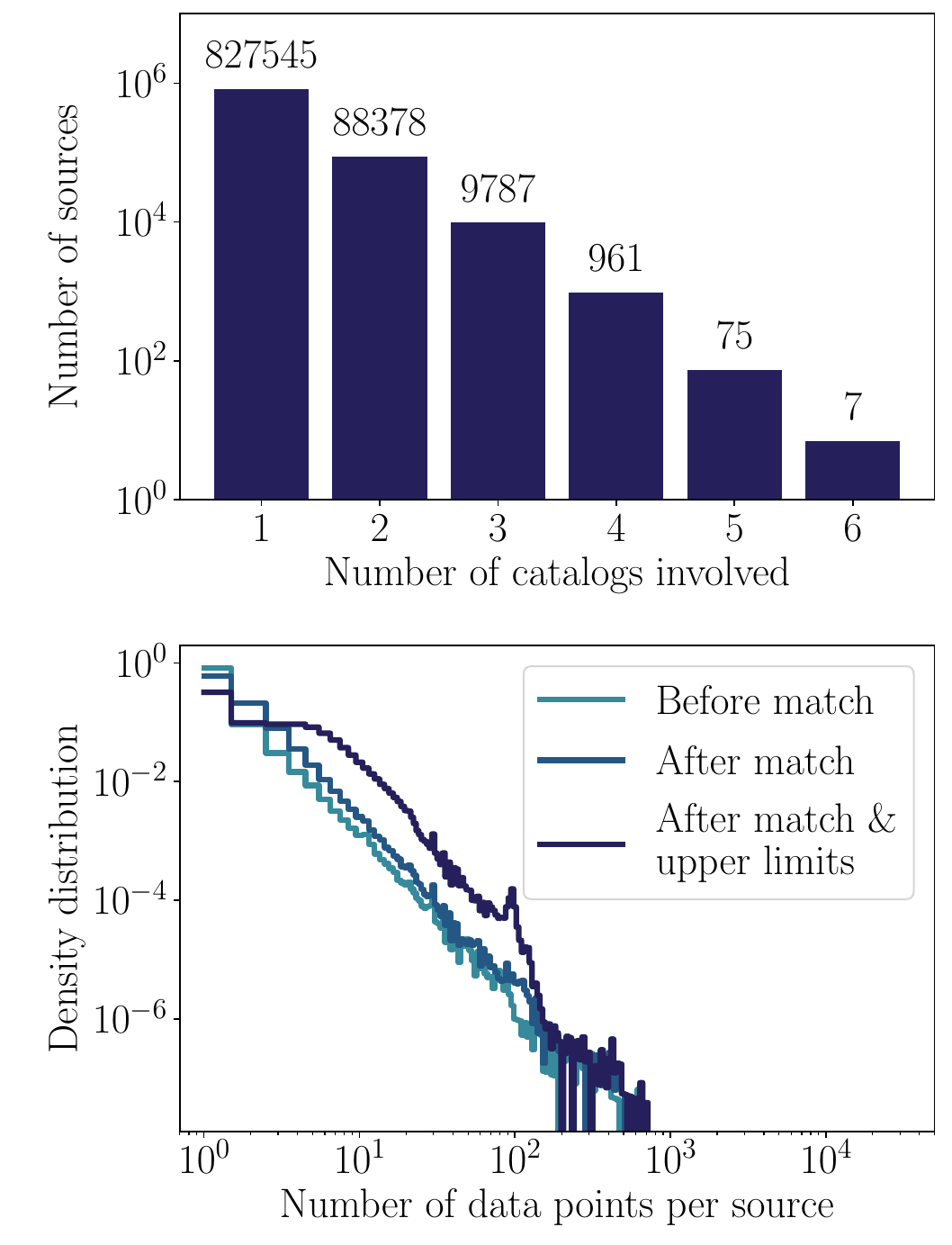}
    \caption{Illustration of the gain in information on the long-term evolution of X-ray sources, obtained thanks to the cross-matching \& upper-limits. \textit{
    Top panel:} Distribution of the number of catalogs involved in each multi-catalog source. The majority of the sources only have data for one catalog, but for the remaining 15\% at least two catalogs are involved. Despite using 7 catalogs, no source was detected in all of them (mostly due to the very constraining sky coverage of the eFEDS catalogs). \textit{Bottom panel:} Density distribution of the number of data points per source, before the cross-match in light blue, after the match in blue, and after taking into account upper limits in dark blue. Both the cross-match and the use of upper limits allows us to increase the number of data points per source, namely, skew this density distribution to the right.}
    \label{fig:CatalogsAndDetsPerSource}
\end{figure}

The cross-matched catalog consists of 926 753 multi-catalog sources, to be compared with the initial 1 258 420 single-catalog sources before the cross-match. Because of the sparse X-ray coverage of the sky (see Fig.\,\ref{fig:Coverage} for a sky map of the final catalog), most of the final sources only contain data from one catalog, but the remaining 15\% of the final sources (99 208) show multi-catalog data (see top panel in Fig.\,\ref{fig:CatalogsAndDetsPerSource}). The catalog-wise matching properties in terms of number and typical offsets are summarized in Table\,\ref{tab:MatchStats}, and the distribution of number of catalogs per cross-matched source is shown in Fig.\,\ref{fig:CatalogsAndDetsPerSource}.

The underlying goal of this multi-catalog method was to increase the number of data points available per source, in order to be able to better estimate the underlying object's variability. The catalog cross-matching allowed us to increase the average number of detections per source from 1.55 to 1.75. The use of upper limits 
allowed us to further increase the average number of data points (detections and upper limits combined) from 1.75 to 5.0 (see precise statistics in the next Section). The precise density distributions of the final number of data points per source is available in Fig.\,\ref{fig:CatalogsAndDetsPerSource}. In particular, the number of sources with only one detection (i.e., for which estimating the variability is impossible) went down from 839 361 to 675 829 thanks to the instrument cross-matching, and is further reduced to 302 252 once upper limits are taken into account. For sources which already had several available data points, the cross-matching allows us to improve the temporal coverage of the source, either by diminishing the average time between two consecutive data points, or by increasing the total coverage (i.e., time between the first and last available data points).

\begin{table*}[]
\centering
\begin{tabular}{c||c|c|c|c|c|c|c}
Cross-match & Chandra & Swift & eFEDS & XMM Slew & ROSAT Survey & ROSAT Pointed & XMM Stacked\\
&&&&&&& (without pointed)\\
\hline \hline
\rule{0pt}{1.\normalbaselineskip}XMM Pointed 
&48106&27710&1364&1368&1408&6294&N/A\\
&1.4"&2.6"&3.6"&5.4"&13.8"&9.9"\\\hline
\rule{0pt}{1.\normalbaselineskip}Chandra 
&&10055&177&558&619&2472&1537\\
&&2.3"&3.2"&5.8"&13.3"&11.6"&1.1"\\\hline
\rule{0pt}{1.\normalbaselineskip}Swift 
&&&281&3345&4114&3992&343\\
&&&3.8"&5.6"&12.8"&11.5"&2.3"\\\hline
\rule{0pt}{1.\normalbaselineskip}eFEDS 
&&&&52&148&1&34\\
&&&&5.9"&14.3"&20.7"&3.2"\\\hline
\rule{0pt}{1.\normalbaselineskip}XMM Slew 
&&&&&4690&1721&15\\
&&&&&12.9"&17.8"&5.2"\\\hline
\rule{0pt}{1.\normalbaselineskip}ROSAT Survey
&&&&&&3865&14\\
&&&&&&31.5"&14.3"\\\hline
\rule{0pt}{1.\normalbaselineskip}ROSAT Pointed 
&&&&&&&77\\
&&&&&&&10.2"\\\hline
\end{tabular}
\caption{Final two-by-two cross match statistics of our multi-instrument catalog. For each combination of catalogs, we show the number of final multi-instrument sources involving both the catalogs, as well as the median angular distance between these sources. As a reminder, we did not compute the \xmm~pointed to \xmm~stacked cross-correlation, as this work was already performed and manually screened in the elaboration of the \xmm~stacked catalog.}
\label{tab:MatchStats}
\end{table*}

\subsubsection{Upper limits statistics}
We called RapidXMM on the 586 483 multi-instrument sources that lie in the \xmm~MOC -- out of those, 116~926 are not 4XMM DR11 sources. Half of these (65~939) are faint \chandra~sources, and the rest are either \xmm~Stacked detections with no clean association in the normal catalog (31~628), \swift~stacked detections, or some \xmm~slew transients or unflagged spurious detection (mostly in extended sources for which the \xmm~slew extent is falsely zero due to low counts). 

\begin{figure}[h]
    \centering
    \includegraphics[width=\linewidth]{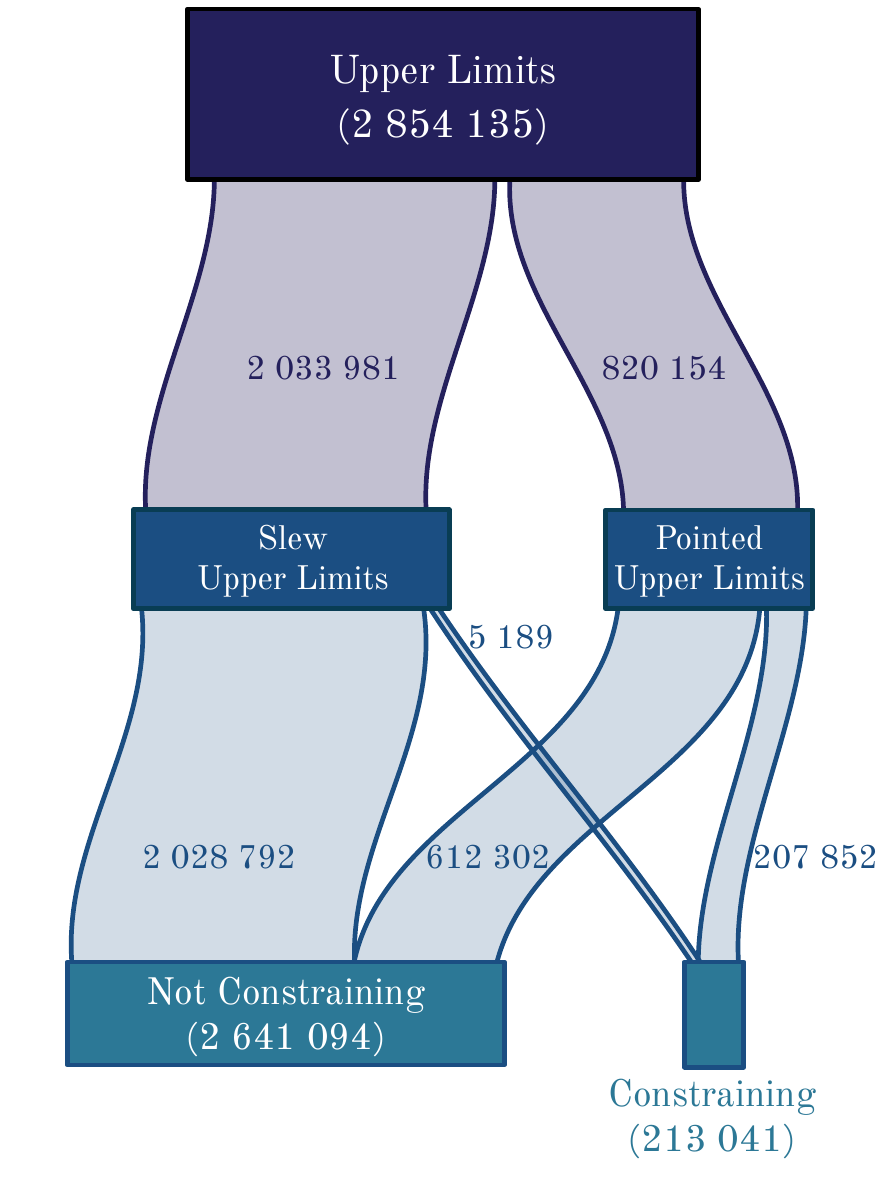}
    \caption{Statistics for the 2 854 135 RapidXMM upper limits on multi-instruments sources in the 4XMM DR11 MOC. These combine the three EPIC instruments, and are 0.2--12 keV flux $3\sigma$ upper limits. An upper limit is considered constraining if it is lower than the lowest flux value of the corresponding multi-instrument source. Most upper limits are from the slews of the catalog, although these are seldom constraining.}
    \label{fig:UpperLimits}
\end{figure}

The statistics of the resulting upper limits are shown in detail in Fig.\,\ref{fig:UpperLimits}. We retrieved 2~854~135 upper limits, 70\% being \xmm~slew upper limits and 30\% being for pointed observations. The overwhelming majority (92\%) of these upper limits are not constraining, in the sense that they are higher than the lowest recorded flux of the corresponding multi-instrument source. However, for 213~041 upper limits (corresponding to 63~795 individual multi-instrument sources), they are indeed constraining, thus allowing us to improve our constraint on the variability of the underlying objects. Among these sources, 13~497 do not correspond to either an \xmm~pointed or stacked source, meaning that a multi-instrument approach was necessary in constraining the variability of the underlying objects.

\begin{figure}[h]
    \centering
    \includegraphics[width=\linewidth]{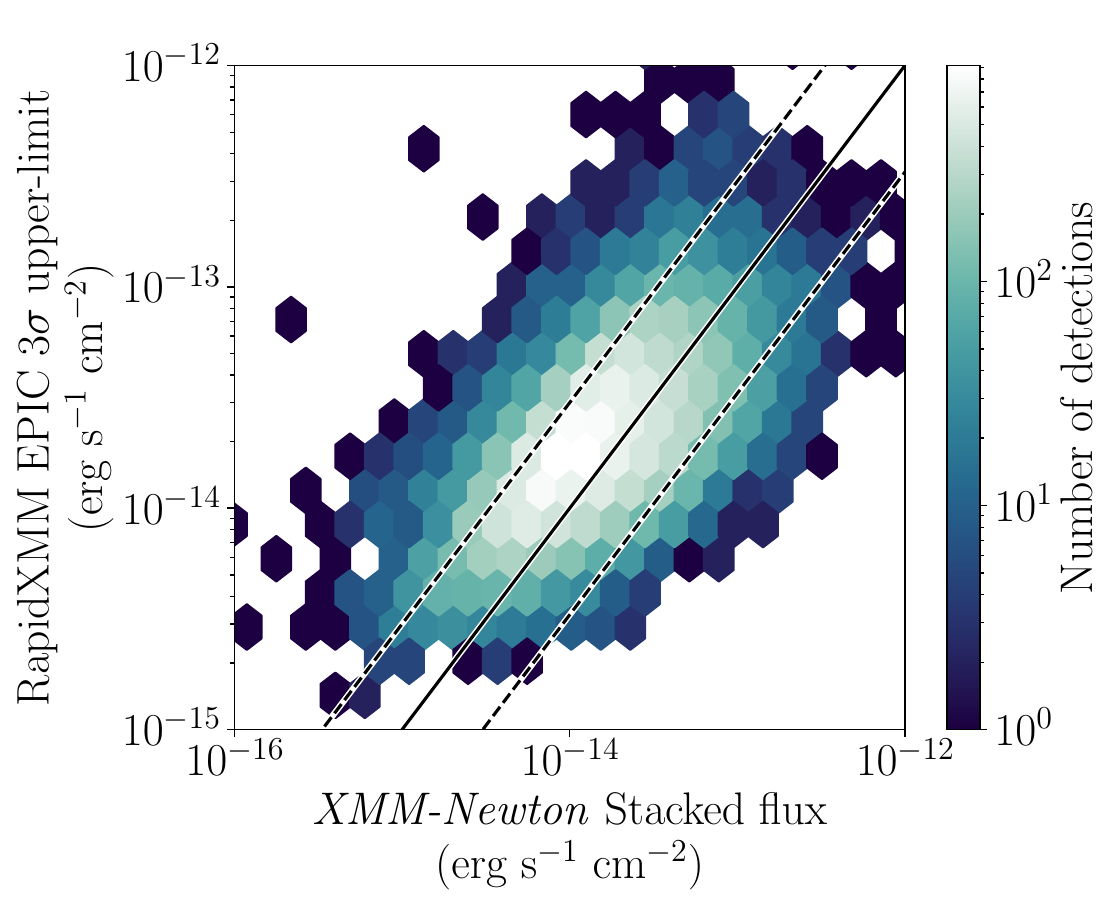}
    \caption{Comparison between the RapidXMM $3\sigma$ 0.2-12 keV flux upper limits, and the corresponding \xmm~stacked 0.2--12 keV flux detections. The black line shows a one-to-one behavior, and the dashed black lines show a departure by a factor of three from this behavior.}
    \label{fig:Stacked_vs_Rapid}
\end{figure}

We chose not to use RapidXMM upper limits in the case where a flux value is available from the \xmm~stacked catalog, which provides measurements in all covering \xmm~observations. This was justified by the additional manual screening that the \xmm~stacked catalog went through. However, as a side result, we were able to assess the quality of the RapidXMM upper limits by comparing them to the simultaneous \xmm~stacked detections, which underwent several additional steps of screening. The resulting comparison between the 22 161 relevant detections is shown in Fig.\,\ref{fig:Stacked_vs_Rapid}. Overall, the majority (82\%) of the RapidXMM $3\sigma$ upper limits are within a factor of three of the corresponding \xmm~stacked detection. Once the \xmm~stacked flux error bars are taken into account, this fraction goes up to 99\%, demonstrating coherence between the two methods. In particular, this confirms the quality of the RapidXMM flux constraints in the case where no \xmm~stacked source is present, that is, transients that were bright in another catalog.

\subsubsection{Variability statistics}
After performing both the catalog cross-correlation and \xmm~upper limits computation, we obtain a large multi-instrument X-ray archival catalog. While such a tool can have various applications for data mining endeavours, systematically exploiting this catalog is beyond the scope of this work. However, we are particularly interested in one information, the long-term variability of sources. Among the various ways to define the variability of an object, we chose to use the pessimistic flux variability amplitude: 
\begin{equation}
    \label{eq:variabilityDef}
    V = \frac{max(F_{\rm low})}{min(F_{\rm up}, UL)}
\end{equation}
where $F_{\rm up} = F + \sigma^{+}$ corresponds to the flux 1$\sigma$ upper value when there is a detection (with $F$ the fluxes and $\sigma^{+}$ the $1\sigma$ positive flux error), $UL$ corresponds to the 3$\sigma$ upper limit when there is no detection (as obtained through RapidXMM), and $F_{\rm low}$ corresponds to the flux lower value in the case of detection, precisely given by $F_{\rm low} =  max(F-\sigma^{-}, 0)$, with $\sigma^{-}$ as the $1\sigma$ flux negative error. Such a definition of $F_{\rm low}$ is meant to avoid it being negative number, as this would contaminate the value of $V$. If a flux measurement is unconstrained (i.e., $F-\sigma^{-}\leq0$), then this point is essentially ignored in the computation of $max(F_{\rm low})$ if there are other well-constrained data points. Using this definition of the variability $V$ allows us to estimate simultaneously the amplitude and significance of the variability. If $V<1$, it means that the various data points are consistent at the $1\sigma$ level, namely, the source appears constant over time. However, if $V>1$, its value gives a lower limit on the actual variability of the underlying physical object. It is important to note here that the variability value we measure is always at best a lower limit of the actual physical variability, due to the sparsity of the X-ray coverage.

\begin{figure*}[h]
    \centering
    \includegraphics[width=\linewidth]{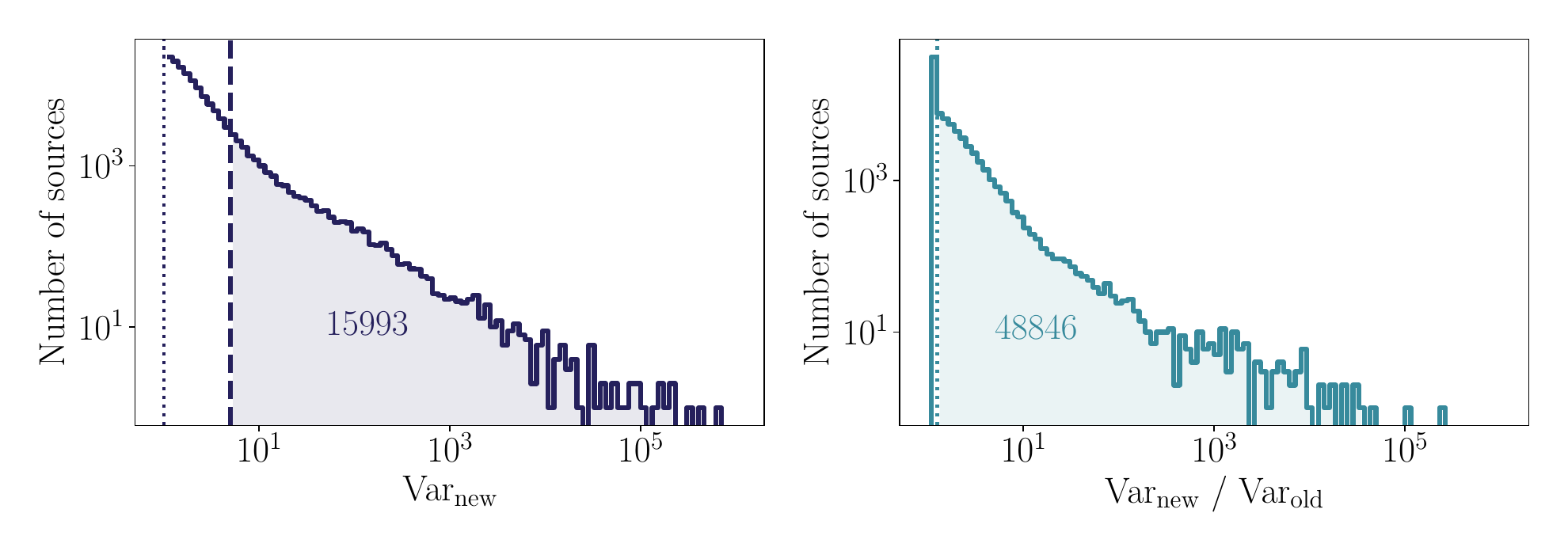}
    \caption{Illustration of the long-term X-ray variability revealed by our method. \textit{
    Left panel}: Distribution of the variability for the multi-instrument sources, including the \xmm~upper limits. We only show sources consistent with being variable (i.e., Var$_{\rm new}$>1, on the right of the vertical dotted line). The vertical dashed line shows the arbitrary limit for what we consider as significant variability (i.e., pessimistic amplitude above 5). Out of the $\sim$135 000 sources with Var$_{\rm new}$>1, only $\sim$16 000 have Var$_{\rm new}$>5. \textit{Right panel}: Distribution of improvement of variability between all the initial single-catalog sources for which a variability estimate was available, and the final multi-instrument source. The vertical dotted line signifies the limit between single-catalog sources for which the new variability is larger than the prior estimate ($\sim$49 000 sources out of $\sim$95 000), and the ones where the new method does not improve the variability estimate ($\sim$46 000).}
    \label{fig:VarGainDistrib}
\end{figure*}

Since our cross-matching and upper limits method was meant to improve our constraints on the variability of X-ray objects, we can now assess the effectiveness of our method using this definition of the variability. As was explained in the previous sub-section, our method decreased the number of sources with one data point only, namely, increased the number of sources for which the variability can be estimated. The distribution of variability for the multi-instrument sources is shown in detail in Fig.\,\ref{fig:VarGainDistrib}, as well as the gain in variability made using our method. Before the cross-matching, there were 74~030 single-catalog sources with a variability estimate over 1 (out of the 207~966 where the estimate was available, and the 1~258~420 total single-catalog sources), and 4~622 with a variability larger than 5. Thanks to our method, out of the resulting 926~753 multi-instrument sources, 618~816 have a variability estimate, which is above 1 for 134 997 multi-catalog sources and above 5 for 15~993 of them. The fraction of variable sources compared to the complete catalog is thus increased from 5\% to 15\% using our method. The fraction of significantly variable sources ($V>5$) is also increased from 0.3\% to 1.7\%. The arithmetic mean gain of variability from the single-catalog sources to the multi-catalog sources is $\sim$10 (see Fig.\,\ref{fig:VarGainDistrib}), although this is mostly driven by few outlying sources with very large gains. The geometric mean of the variability gain (less contaminated by outliers) is $\sim$1.4. This means that our method is successful in improving the constraint on the X-ray variability of archival sources.

\section{The STONKS algorithm}
\label{sec:STONKS}
\subsection{Motivation and possible implementation within the \textit{XMM-Newton} pipeline}

This section presents a possible implementation of our work in the \textit{XMM-Newton} pipeline. This is of course subject to modifications if and when it is to be actually implemented in the data stream.

Currently the new \xmm~observations follow a 1-year proprietary period for non-Heritage data during which the data are only available to the P.I. of the corresponding \xmm~proposal (see the \textit{XMM-Newton} Announcement of Opportunity\footnote{\href{https://xmm-tools.cosmos.esa.int/external/xmm_user_support/documentation/AOpolicy/Policies_Procedures.pdf}{https://xmm-tools.cosmos.esa.int/external/xmm\_user\_support/\\documentation/AOpolicy/Policies\_Procedures.pdf}} for more details). If a transient event was to take place serendipitously within the field of view, and the P.I. failed to detect and identify it, this proprietary period means that any later identification and follow-up processes would take place more than a year after the initial detection. This entails a loss of most of the valuable early-time physical information which could have been gathered if the transient had been immediately detected. For this purpose, we have developed the "Search for Transient Objects in New detections using Known Sources" algorithm (STONKS).

The suggested framework of STONKS is as follows. Once the \xmm~observational data have been downloaded from the satellite, they go through an automatic processing and source-detection pipeline. As part of the ACDS pipeline, the EPIC summary source list could then be provided to STONKS, in order to check for long-term variability. This would automatically generate a PDF file for each alert in the field of view. This file can be sent to the P.I. of the observation, as part of the PPS products. Additionally, at this point, the pipeline products are checked manually by an \xmm~scientist \citep[e.g.,][]{watson_xmm-newton_2009} -- we suggest that the alerts are also checked by the \xmm~scientist, who will then validate them. After validation, they will be uploaded to a database hosted at IRAP. If the P.I. expressed their agreement  and the source is serendipitous, the alerts are then made available on a public web service. 

\begin{figure*}[h]
    \centering
    \includegraphics[width=\textwidth]{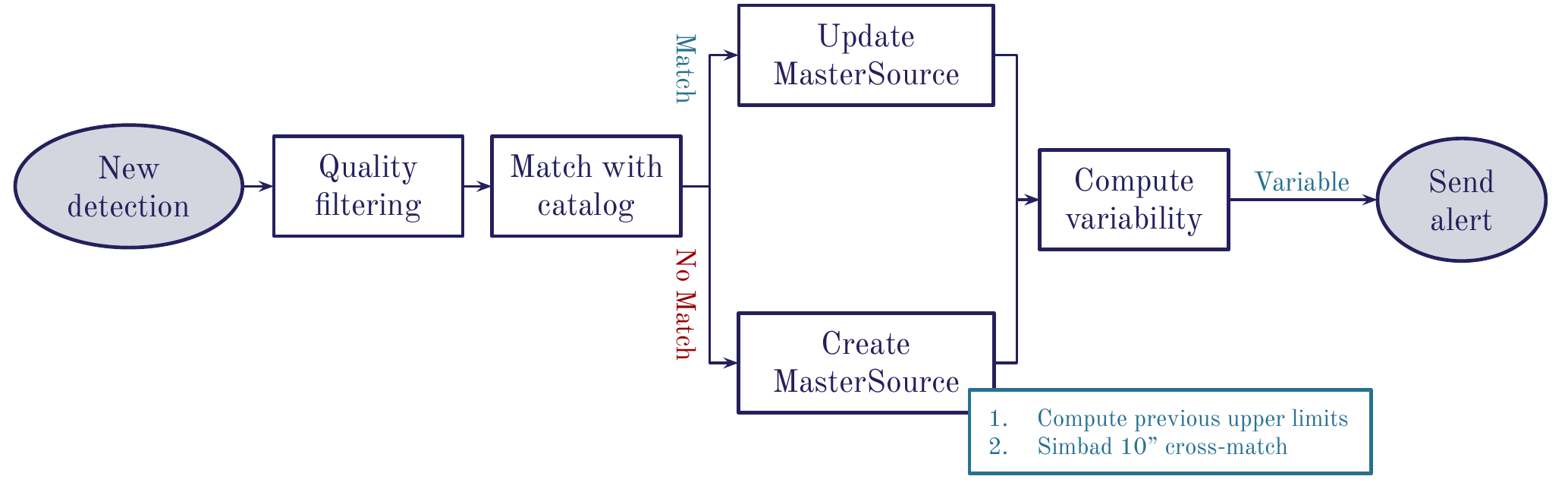}
    \caption{Schematic representation of the workflow of STONKS on a given new \xmm~detection. The main differences in treatment arise from the result of the cross-match with the archival multi-instrument catalog. A detection is considered "variable" if the associated multi-instrument source (called "MasterSource" here) has a long-term variability larger than five, as defined in \ref{eq:variabilityDef}.}
    \label{fig:Workflow}
\end{figure*}

The suggested workflow that would be then followed by each detection is presented in Fig.\,\ref{fig:Workflow}. The new detections would be filtered based on their quality criteria. To be more precise, we require the extent likelihood to be below 6 (to keep only point-like sources), and the detection likelihood to be over 10 ($\sim4\sigma$) in all EPIC instruments for which the source is in the field of view in order to retain the most reliable sources. Indeed, after initial testing we found that detections for which some instruments had low detection likelihoods but other instruments had sufficient detection likelihood tended to be dominated by spurious detections and instrumental effects. The remaining clean detections would then be first cross-matched with the archival multi-catalog sources, using the 3$\sigma$ position error, and the same ambiguity-solving framework as was used when building the catalog. If the ambiguity cannot be lifted, we cannot safely confirm any long-term variability, so the process stops at this stage. Otherwise, there are two situations: either the source is new and does not match any of the archival sources, in which case the previous possible upper limits would be computed by calling RapidXMM on the source's position, and a 10" Simbad cross-match performed using the \texttt{astroquery} package \citep{ginsburg_astroquery_2019}. If the source matches the archival catalog without ambiguity (or if this ambiguity is solvable), then the new detection can be added to the multi-catalog source's history. For both cases, STONKS would then assess the new long-term variability of the source, given this new information. If the multi-catalog source, with the added detection, is overall variable with a pessimistic variability amplitude over five (as was defined in Eq. \ref{eq:variabilityDef}), a variability alert associated with the detection would be raised.

The output would be presented in the form of a PDF file, with four panels (see examples in Fig. \ref{fig:SpectralErrorAlert} to Fig. \ref{fig:AbsorbedAGN}). The first contains the long-term multi-instrument light curve, including upper limits, extrapolated to the 0.1--12 keV band. The second panel contains the band photometry of each detection, allowing us to assess spectral variability in the source, or spurious flux variability due to extreme softness or hardness of the source (see Sect. \ref{subsec:FluxCal}). The third panel contains a 2'$\times$2' optical image of the source from the Digital Sky Survey \citep{lasker_palomar--st_1996}, queried using the \texttt{astroquery} package. Finally, the fourth panel contains details about the observation itself (observation identifier, date, name of the target), about the detection (identifier of the detection in the observation, position and position error, off-axis angle and detection likelihoods in the three EPIC instruments), and about the associated multi-catalog source (type of alert, long-term and short-term variability, and SIMBAD classification if the source was already known). There are four possible types of alerts:
\begin{itemize}
    \item "High-flux state" if the new detection is the brightest historical state of the multi-catalog source;
    \item "Low-flux state" if it is the lowest historical state (including lower than past \xmm~upper limits);
    \item "First-detection" if this is the first time the source is detected, with prior upper limits being constraining. This is technically similar to "High Flux State", but might be more sensitive to spurious detections, hence the separate category;
    \item "Past-variability" in the case where the new detection is between the brightest and dimmest historical states of the multi-catalog source and this source has shown variability in the past.
\end{itemize} 
Finally, we added a warning regarding the spectral assumption. This warning is raised if any of the detections of the source (including the new detection) have a spectral hardness that falls into the 10\% hardest or softest detections of its respective catalogs. This could potentially mean that the variability is mis-estimated. The corresponding thresholds are presented in Table \ref{tab:EnergyConvFactor}. Various examples of serendipitous alerts are available in Sect. \ref{sec:AppTesting}. The precise format of the alert PDF file is of course subject to change, depending on the various feedbacks from the \textit{XMM-Newton} scientists and the community, once the service is operational.

We recommend the alert would then be returned to the \xmm~scientist for manual screening -- this would expand the screener's task, but the expected number of alerts is reasonably low (see Sect. \ref{sec:Testing}). Alerts that are not spurious could then be shared using one of the standard community mechanisms. We also intend to upload the alerts as a JSON file to a database hosted at IRAP, that would then be displayed on a publicly available web service (the precise details for this service; for instance: a possible notification system, are yet to be determined). 
STONKS is currently publicly available through a REST API \footnote{\href{https://xcatdb.unistra.fr/stonks/}{https://xcatdb.unistra.fr/stonks/}} which takes a \textit{XMM-Newton} EPIC observation source list as an input (POST request) and returns a tarball with all the PDF corresponding the detected variability. The service can be connected either from a WEB page or through clients such as CURL.

\subsection{Testing}
\label{sec:Testing}
To assess the validity of our method, we simulated the behavior of the alert system over archival \xmm~data. We ran STONKS on the 483 observations from 2021 for which the observing mode allows us to observe serendipitous sources, checking variability for 12~584 detections, leading to 315 individual alerts (alert rate of $\sim2.5\%$ among all the detections). The various statistics of these alerts are represented in Fig. \ref{fig:StatAlerts}.

\begin{figure}[h]
    \centering
    \includegraphics[width=\linewidth]{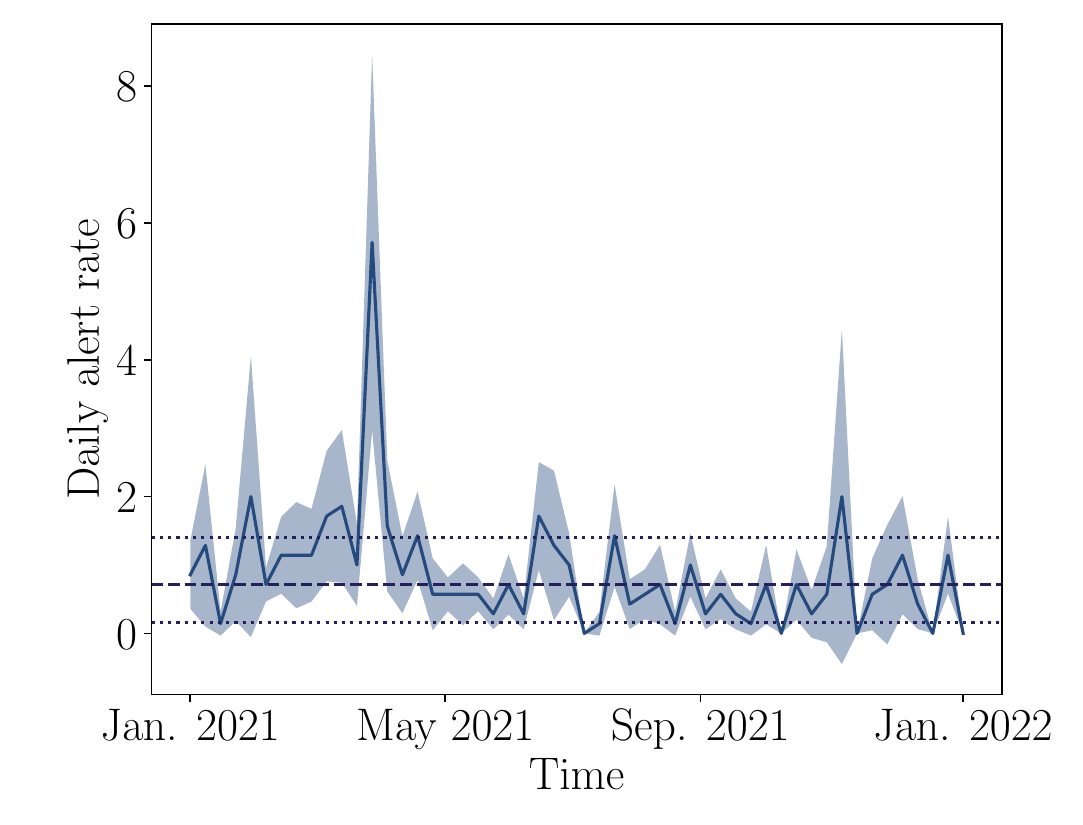}
    \caption{Daily alert rate computed on a weekly average. The envelope corresponds to the standard deviation of this daily rate over each week. The dashed and dotted lines correspond to the yearly median and $1\sigma$ errors on the rate of $0.7^{+0.7}_{-0.5}$ alerts per day. The large peak at the end of March corresponds to a set of several consecutives observations of Sgr A*, simultaneous to GRAVITY exposures -- the Galactic center is particularly rich in X-ray transient events, either stellar flares or bursts from X-ray binaries.}
    \label{fig:PipelineTest}
\end{figure}

The evolution of the resulting daily alert rate over the testing run can be seen in Fig.\,\ref{fig:PipelineTest}, with a daily rate of $0.7^{+0.7}_{-0.5}$ alerts per day. 
The standard deviation of this daily rate is quite large, as the number of alerts in a given observation is highly dependent on the specific targeted field of view (e.g., the Galactic center is rich in transients).

Out of these 315 alerts, 53 were the target of the observation, while 262 were serendipitous sources. Since the idea behind STONKS is to detect previously unknown transient events, this large fraction ($\sim80$\%) of serendipitous alerts is encouraging. Even for the target of the observation, an assessment of the long-term variability might be useful for the P.I. of the observation. 

Among the 315 alerts, about 40\% were linked to past variability events (138), the remaining three categories being about evenly distributed (68 "low-flux state" alerts, 52 "high-flux state" alerts, and 57 "first-detection" alerts). Overall, the target sources have a slightly higher fraction of "past-variability" alerts (28 out of 53) than the serendipitous sources (110 out of 262). This difference is mainly driven by the much larger fraction of "high-flux state" and "first-detection" alerts for serendipitous sources -- this is expected for serendipitous transients happening in the field of view. Seven "first-detection" alerts were sent for targets of an observation, showing two limitations of our method. For four of these alerts, they were linked to a high proper motion object (in this case the M dwarf CN Leo): since our matching methods and upper limit computation work is based on a fixed sky position, high proper motion objects will naturally lead to spurious transient alerts. Correcting this issue would require retrieving the proper motions of the sources in and near the field of view, and compensating it in the various position-dependent steps of our algorithms, which is beyond the scope of our approach. The three remaining alerts were linked to a new TDE detected by eROSITA \citep[eRASSt J045650.3-203750, e.g.,][]{malyali_erasst_2023,liu_deciphering_2023}. While it is reassuring to trigger an alert on a TDE in the field of view, the fact that three alerts were sent out for the same object is due to the fact that STONKS does not update its archival database on new detections. This is meant to avoid spurious detections contaminating the catalog before they are filtered out by manual screening. However, it will lead to multiple alerts being sent out in the case where the source was detected several times since the last data release of the catalogs. This also prevents the detection of variability between two observations of a given data release. This precise approach might be subject to change in ulterior versions of STONKS, with for instance the inclusion of detections from the same data release (after manual screening), with an additional warning about them.

Using the 10" cross-match with Simbad, we retrieve classification for a fraction of the alerts (113 out of 315 -- see Fig. \ref{fig:StatAlerts}). Out of these, 30 correspond to X-ray binaries, 36 to stellar objects, and 47 to galaxies or AGNs. For the remaining alerts, 63 do not have a specific classification in Simbad, which usually indicates that they are part of a large scale catalog (e.g., "X-ray source", as part of a X-ray telescope catalog with no individual classification). For 139 alerts, they are not at all in Simbad -- manual inspection indicates that these are mostly stellar objects. Almost all alerts corresponding to first detections (i.e., using past upper limits) have no Simbad counterpart.

Out of the 315 alerts, the contamination rate is estimated after manual screening to be below 20\%. These errors are driven by high proper motion objects, instrumental errors, and more frequently failures of the spectral assumption (as explained in Sect. \ref{subsec:FluxCal}). The false alert rate of $\sim0.6\%$ presented in Sect. \ref{subsec:FluxCal} can be compared to the $\sim2.5\%$ total alert rate per detection we obtained on the 2021 data, confirming the estimated $\sim20\%$ contamination. While it is difficult to avoid these issues in our pipeline, the output alert was designed to help manually identify these possibilities. The second panel, showing the band photometry of each X-ray detection, allows us to roughly compare their corresponding spectra and see if they are compatible, despite the flux estimates showing variability. This can be seen for instance in the spurious alert in Fig.\,\ref{fig:SpectralErrorAlert}: the source being quite hard, the extrapolation between instruments will introduce a bias in the flux estimates, but the spectra are clearly compatible. It is then straight-forward to discard this alert. For the high proper motion objects, the optical view provided in the third panel can allow us to see these objects, as a bright nearby star will appear slightly off-centered from the targeted position. A proper manual screening needs to be performed in order to confidently remove these alerts. Finally, the instrumental errors and spurious detections are hard to exclude in a catalog-oriented approach. Since these alerts will be dealt manually, it will be possible to discard those corresponding to manually flagged spurious detections.

\subsection{Some variable sources found during the testing run}

The idea behind STONKS is to allow us the community to quickly detect X-ray serendipitous transient objects, and follow up on them if relevant
. We show in this Section a (somewhat arbitrary) selection of some variable objects found in the 2021 test run of STONKS. These include a possible TDE candidate, AGNs with long-term or short-term (i.e., over the course of a single observation) spectral variability, a flaring star and new XRB and ULX candidates. 

For each of these sources, we used the EPIC pn data when available, and the MOS data otherwise. We performed the standard procedure from the \xmm~data analysis threads\footnote{\href{https://www.cosmos.esa.int/web/xmm-newton/sas-threads}{https://www.cosmos.esa.int/web/xmm-newton/sas-threads}}, using SAS 19.0.0\footnote{”Users Guide to the \textit{XMM-Newton} Science Analysis System”, Issue 18.0, 2023 (ESA: XMMNewton SOC)} and Xspec \citep{arnaud_xspec_1996} for the spectral fitting.

\subsubsection{4XMM J151509.5+561347: TDE or flaring AGN?}

4XMM J151509.5+561347 showed a soft outburst in August 2021 (ObsID 0891801501), with a variability of a factor $>13$ compared to previous upper limits (see the alert \ref{fig:Tde_Alert}). Its optical counterpart (SDSS J151509.61+561347.3) is classified as a galaxy \citep{ahumada_16th_2020}, with a photometric redshift of 0.33$\pm$0.09. The nearby galaxy, SDSS J151510.27+561344.7, is brighter and has a spectroscopic redshift of 0.16. Using the photometric redshift of 0.33$\pm$0.09, the peak flux value of $\sim (7\pm1)\times10^{-13}$ erg s$^{-1}$ cm$^{-2}$ translates into a luminosity of $2.5^{+2.5}_{-1.5}\times 10^{44}$ erg s$^{-1}$. This type of luminosity can be reached by both high accretion episodes in AGN or bright TDEs at their peak. The soft emission is consistent with both as well, however the spectrum (see Fig.\,\ref{fig:TDEspectrum}) is better explained by an absorbed powerlaw ($\chi^{2}$/DoF = 24.5/18, $\Gamma=2.7\pm0.4$) than by an absorbed black body ($\chi^{2}$/DoF = 75/18, $k_{B}T=173\pm8$ eV). It is hard to clearly discriminate between these two possiblities based on the spectral shape only. Ideally, a timely X-ray and / or optical follow-up would have allowed us to assess the presence of either AGN of TDE emission, based on the spectral-timing properties of the emission after the peak (e.g., a $\propto t^{-5/3}$ decay over a few months for a TDE, compared to the red noise expected in an AGN).


\subsubsection{4XMM J000532.8+200717: a quasar with variable photon-index}
4XMM J000532.8+200717 is a quasar at $z=0.119$ \citep{caccianiga_xmm-newton_2008} that showed a significant long-term spectral variability over the 20 years of available X-ray data (see alert Fig.\,\ref{fig:QuasarSpecVar_alert}). It underwent an episode of high emission in the late 2000s, with noticeable Swift variability of about an order of magnitude (between luminosities of $\sim10^{43}$ to $\sim10^{44}$ erg s$^{-1}$). It is noticeably harder at the peak than in quiescence (see Fig.\,\ref{fig:Quasar_SpecVar_Spectra}). The peak spectrum is consistent with an intrinsically absorbed power law ($N_{\rm H}^{\rm Peak}=(1.0\pm0.5)\times10^{20}$ cm$^{-2}$ and $\Gamma^{\rm Peak}=3.2\pm0.1$), with a much softer photon index in the low state and consistent intrinsic absorption ($N_{\rm H}^{\rm Low}=(5\pm3)\times10^{20}$ cm$^{-2}$, and $\Gamma^{\rm Low}=5.2\pm0.6$). This change is further confirmed by the fact that freezing the photon index at the peak value and fitting only the normalization and absorption on the low state significantly worsens the fit statistics, from $\chi^2/$DoF=30/17 to 52/18.

\subsubsection{4XMM J053231.0+170504: a typical stellar flare}
4XMM J053231.0+170504 is a star \citep[TYC 1301-1536-1, from][]{hog_tycho-2_2000} that showed significant X-ray variability by a factor $\sim6$ between two \textit{XMM-Newton} observations two years apart (see Fig. \ref{fig:StellarFlare_alert}). Its long-term variability is in fact a consequence of the large short-term flare it underwent during the second \textit{XMM-Newton} observation, which has an impact on the observation-averaged flux (see Fig. \ref{fig:StellarFlareLightcurve}). Such X-ray flares, of amplitude $\sim5$ and timescale $\sim2$ ks, are expected from active stars \citep[e.g.,][]{benz_physical_2010}.

\subsubsection{4XMM J000532.8+200717:  Quasar with variable photon-index}
4XMM J000532.8+200717 is a quasar at $z=0.119$ \citep{caccianiga_xmm-newton_2008} that showed a significant long-term spectral variability over the 20 years of available X-ray data (see alert Fig.\,\ref{fig:QuasarSpecVar_alert}). It underwent an episode of high emission in the late 2000s, with noticeable Swift variability of about an order of magnitude (between luminosities of $\sim10^{43}$ to $\sim10^{44}$ erg s$^{-1}$). It is noticeably harder at the peak than in quiescence (see Fig.\,\ref{fig:Quasar_SpecVar_Spectra}). The peak spectrum is consistent with an intrinsically absorbed power law ($N_{\rm H}^{\rm Peak}=(1.0\pm0.5)\times10^{20}$ cm$^{-2}$ and $\Gamma^{\rm Peak}=3.2\pm0.1$), with a much softer photon index in the low state and consistent intrinsic absorption ($N_{\rm H}^{\rm Low}=(5\pm3)\times10^{20}$ cm$^{-2}$, and $\Gamma^{\rm Low}=5.2\pm0.6$). This change is further confirmed by the fact that freezing the photon index at the peak value and fitting only the normalization and absorption on the low state significantly worsens the fit statistics, from $\chi^2/$DoF=30/17 to 52/18.

\subsubsection{4XMM J081909.2+703928: Possibly misclassified ULX candidate}

4XMM J081909.2+703928 is a hard X-ray source, appearing in the outskirsts of the dwarf galaxy Holmberg II. It showed large variability over the 20 years of available X-ray data, by a factor of about 300 over short timescales ($\sim$days, see alert in Fig.\,\ref{fig:MisclassifiedULX}). It is part of the NuSTAR hard X-ray sources catalog \citep{zappacosta_nustar_2018}, and an optical spectral follow-up for this study assessed a redshift of $z$=1.27, thus making this source an AGN candidate (even blazar candidate, with corresponding variability and lack of spectral change, and peak \textit{Swift} luminosity of $\sim10^{46}$ erg s$^{-1}$). However, the optical counterpart to this source is extremely dim, not even visible in the PanSTARRs survey, meaning that the initial redshift estimate is most likely spurious. 
The absence of an optical counterpart also excludes the blazar interpretation, which should be bright in optical light as well, seeing as there is no sign of absorption in the X-ray spectrum (see next paragraph).

Ignoring the pre-existing redshift estimate, another possibility is that the source is in the periphery of Holmberg II, and not a background source. This could be strengthened by the presence of a faint UV detection in the \xmm~Optical Monitor (XMMOM J081909.2+703929, with a UVW1 flux of $\sim10^{-17}$~erg~s$^{-1}$~cm$^{-2}$~\AA$^{-1}$), without optical counterpart, which could correspond to a faint star cluster. Assuming it is located at the same distance as Holmberg II \citep[i.e., 3.39 Mpc,][]{karachentsev_m_2002}, the luminosities range from $10^{37}$ up to $\sim 3\times10^{39}$ erg s$^{-1}$, which is consistent with high-luminosity episodes of an X-ray binary, even reaching ULX-levels of luminosity. The spectrum of a high luminosity episode, for the observation that triggered the alert (ObsID 0864550401) is better fitted by an unabsorbed dual component powerlaw and black body than by a simple unabsorbed powerlaw ($\chi^{2}$/DoF of 37/31 compared to 65/33), as is shown in Fig.\,\ref{fig:MisclassifiedULXspectrum}. Such a double component spectrum is characteristic of ULXs and X-ray binaries \citep[e.g.,][]{koliopanos_ulx_2017}, and less likely for blazars which are in most cases well-fitted by a single powerlaw component. This tends to support the idea that this source has been misclassified as a background AGN, and is in fact a possible candidate ULX (or at least X-ray binary) in the outskirts of Holmberg II.

\subsubsection{4XMM J013650.6+154011: New candidate XRB}
4XMM J013650.6+154011 showed alternating episodes of activity and quiescence over the 20 years of archival data (see alert Fig.\,\ref{fig:NewXRB}. It displayed variability by a factor $\sim10$ on timescales of a few days to a few weeks. This variability was mostly caught by \swift~and \chandra, making any spectral conclusion difficult. Its faint optical counterpart (SDSS J013650.65+154011.3, AB magnitude in the SDSS r band of 20.8), combined with the timescales and amplitude of variability, supports the interpretation of an X-ray binary. This is further confirmed by the peak spectrum, from the observation that triggered the alert (ObsID 0864270101), which is consistent with an absorbed double component emission with a powerlaw and a black body ($N_{\rm H}=6.4^{+4.5}_{-3.7}\times10^{21}$ cm$^{-2}$, $\Gamma=6.0\pm3.0$, $k_{b}T=0.66^{+0.19}_{-0.13}$ keV, $\chi^{2}/$DoF = 32/32, see Fig.\,\ref{fig:XRBspectrum}), which is typical of X-ray binaries. The other interpretation for such variability would be a blazar, which would have a brighter optical counterpart and is thus excluded.




\subsubsection{4XMM\,J023228.8+202349: Short-term variable absorbed AGN}

\begin{figure}
    \centering
    \includegraphics[width=\columnwidth]{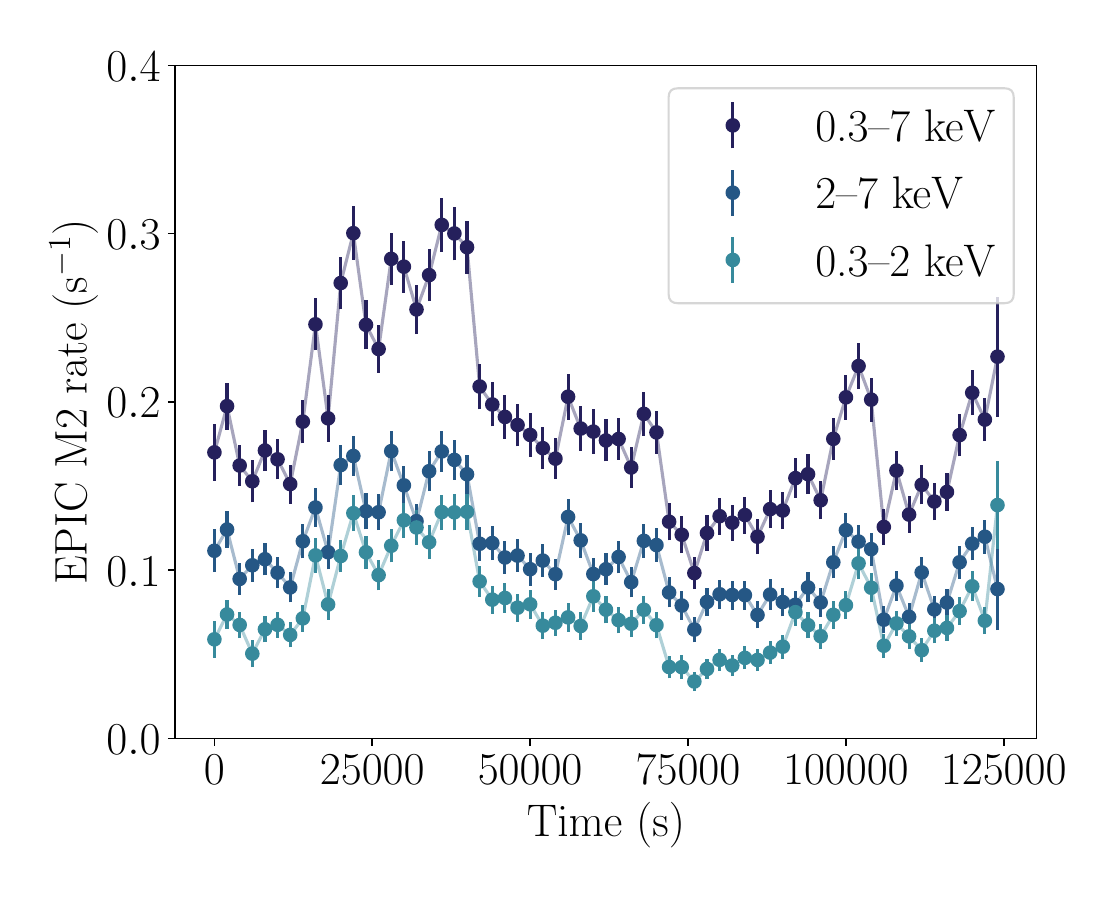}
    \caption{EPIC MOS2 lightcurves of 4XMM J023228.8+202349 (ObsID 0810821801), binned at 2ks. The soft (0.3--2 keV) and hard (2--7 keV) emission evolve in a similar way, meaning that the change is not due to absorption (which would impact more significantly on the soft emission) but is intrinsic to the powerlaw component. A slight $\sim$1ks lag is visible between the soft and hard emission.}
    \label{fig:AbsorbedAGN_lc}
\end{figure}

4XMM J023228.8+202349 is a hard source showing variability by a factor of $\sim$10 over timescales of a few days (see alert in Fig.\,\ref{fig:AbsorbedAGN}). It is part of the NuSTAR serendipitous catalog as well \citep{zappacosta_nustar_2018}, that identified its optical counterpart as a broad-line galaxy at $z=0.029$. The source, in the three available observations, is well fitted with a power law and ionized absorber and a reflection feature (\texttt{TBabs*zxipcf*(zgauss+relxilllp))}. The brightest \xmm~observation, which triggered the alert, is short-term variable as well. The EPIC MOS2 lightcurves can be seen in Fig.\,\ref{fig:AbsorbedAGN_lc}, in several energy bands. There is no difference in the evolution of the soft (<2 keV) and hard (>2 keV) bands, meaning that the change is not in absorption but in the normalization of the power law. The cross-correlation between the soft and hard bands reveals that the soft emission lags slightly ($\sim0.8\pm0.3$ ks) behind the hard emission (see Fig.\,\ref{fig:AGN_lag}). This lag is consistent with the reflection of the hard X-ray corona, which is also confirmed by the spectrum which contains a reflection component (see Fig.\,\ref{fig:HardIonizedAbs_spectrum}). Assuming a constant height of the corona $h$ for the \texttt{relxilllp} component, we find that $h=5.6\pm1.8~r_{g}$. The main changes between the observations are the norm of the power law from $7\times10^{-5}$ to $9\times10^{-6}$ and the column density from $(0.38\pm0.12)\times10^{22}$~cm$^{-2}$ to $(6.5\pm2.5)\times10^{22}$~cm$^{-2}$. The lag of $\sim0.8\pm0.3$ ks is indicative of a size of $(2.4\pm0.9)\times10^{11}$m. Assuming this size is the corona-disk distance, namely, $\sim h$, we find $r_{g}\approx0.4^{+0.4}_{-0.2}\times10^{11}$m, namely, $M_{BH}\approx2.7^{+2.7}_{-1.3}\times10^{7}M_{\odot}$.

\section{Discussion}
\label{sec:Discuss}
\subsection{Implementation in the \textit{XMM-Newton} pipeline and alert dissemination}
STONKS is designed to automatically search for transients in \xmm~EPIC data and can be used to find them in quasi-real time if run at the time of the automatic pipeline processing. These alerts can then be shared with the P.I. of the observation, and with the community, in order to ensure that no transient is overlooked. In essence, it is the \textit{XMM-Newton} equivalent to the newly implemented transient detector for the \textit{Swift} pipeline \citep{evans_real-time_2023}. Another future possibility for making use of these alerts is to create synergies with data brokers for the \textit{Vera C. Rubin} Observatory, such as Fink \citep{moller_fink_2021}.

\subsection{Main limitations and expected results}

As explained in Sect. \ref{sec:Testing}, 
the main limitations of our method are (in decreasing order based on the contamination rate) the failure of the spectral extrapolation assumption in the case of very hard or very soft sources, the presence of instrumental errors and spurious detections, and high proper motion objects for which astrometry-based matching is not straight forward. These issues can be mitigated by manual screening of the produced alert files. Our bayesian cross-match method was successful in avoiding spurious variability based on wrong associations, as no such alert was triggered in the 2021 test run.

The alert rate obtained from the 2021 test run is expected to be representative of the general rate of alerts raised for transients with a variability of at least a factor 5 detected with \textit{XMM-Newton}. While these variable objects are dominated by usual AGN variability and stellar flares, a number of more exotic sources have already been detected in the test run. Only serendipitously detected sources were presented in Sect. \ref{sec:Testing}, as the philosophy behind STONKS is to detect serendipitous variable objects. However, STONKS also would have raised alerts for some variable targeted objects, among which are two TDE candidates -- eRASSt J045650.3-203750, and 4XMM J011104.7-455843. The fact that STONKS was able to catch these targeted objects means that we would also have caught them if they had been serendipitous detections, confirming the efficiency of STONKS. 

\subsection{Updating the archival catalog}
At the time of publication of this work, some catalogs that have been used are already slightly outdated (for instance for \xmm~by two data releases). However, it is our intention to update regularly the archival catalog in use, in order to better be able to detect new transient events. In particular, the inclusion of the eFEDS catalog was meant as a proof-of-concept that, once the eROSITA data from the first all-sky survey are released, it will easily be taken into consideration for future detections. It should theoretically provide us systematically with one data point for comparison, for each new \xmm~detection -- or a possibly constraining eROSITA prior upper-limits in the case of an absence of match between the catalogs. The similarity between the \xmm~and eFEDS sources in terms of flux (see Fig.\,\ref{fig:FluxvsFlux}) is reassuring for the future transient alerts. Additionally, the upcoming \chandra~and \xmm~slew data releases of all observations after 2014, as well as regularly updated versions of the Living \textit{Swift}-XRT Point Sources catalog \citep[LSXPS,][]{evans_real-time_2023}, will also be taken into account.

\subsection{Data mining the archival catalog}
While the focus of this work has been on quasi-real time transient detection, the archival catalog that was built as a by-product of our method is a goldmine for archival variability study. During its elaboration, we have used several criteria to mine it, looking for specific sources of interest. In particular, it allowed us to find a new transient ultra-luminous X-ray source in NGC 7793 with a candidate pulsation \citep{quintin_new_2021}, and a new candidate source of quasi-periodic eruptions in an optically-detected TDE \citep{quintin_tormunds_2023}. Other long-term variable sources, such as new X-ray TDE candidates, have been found in this archival catalog (Quintin et al., in prep).

Our work has mostly focused on long-term X-ray variability estimation and detection. However, others may make use of this archival multi-instrument X-ray catalog for other purposes. For this reason, the cross-matched catalog is made available on both Zenodo\footnote{\href{https://zenodo.org/doi/10.5281/zenodo.10634292}{https://zenodo.org/doi/10.5281/zenodo.10634292}} and the CDS\footnote{\href{http://cdsweb.u-strasbg.fr/cgi-bin/qcat?J/A+A/}{http://cdsweb.u-strasbg.fr/cgi-bin/qcat?J/A+A/}}. 
These files will be updated with each new version of the archival catalog.

\section{Conclusions}
In this paper, we present a new quasi-real time transient detection system for the \xmm~pipeline, STONKS. The general idea of STONKS is to automatically compare any new \xmm~detection to our archival knowledge of the X-ray sky at this position in order to assess the long-term variability of the underlying X-ray source. 

It required a first step of collating most available archival X-ray data. We used the \xmm~pointed, slew, and stacked catalogs, the \chandra~catalog, the \swift~point-like sources catalog, the ROSAT survey and pointed catalogs, and finally the eROSITA eFEDS catalog. We used relatively stringent quality criteria in order to avoid spurious detections, and in particular only kept point-like sources. The catalogs were then matched together two by two at first, with ambiguous correlations being dealt with using a Bayesian framework similar to that of \texttt{NWAY} \citep{2018MNRAS.473.4937S}.  The main difference between our method and \texttt{NWAY} is that, at the two-by-two matching phase,
catalogs are considered in a symmetrical way (whereas \texttt{NWAY} is inherently asymmetrical, looking for the best counterpart for each source of the primary catalog in the secondary catalog). The two-by-two correlations are then merged in a multi-instrument catalog, in a conservative manner by refusing any "chain-like" association between more than two catalogs. This provided us with a catalog of 926 753 multi-instrument sources, with 15\% of them containing information from multiple catalogs. In order to be able to compare flux values between instruments with varying energy bandwidth, we need to convert these instrument-specific fluxes to a common band and, more precisely, the largest possible band using these catalogs, 0.1--12 keV. This extrapolation is done using a fix spectral assumption (absorbed power law with $N_{\rm H}=3\times10^{20}$~cm$^{-2 }$ and $\Gamma$=1.7). This assumption is reasonable for most X-ray sources, and is used in the \xmm~catalogs. We estimated the rate of false positives to be about 0.5\% of the total detections and less than $\sim20\%$ of the alerts, corresponding to the spectrally hardest and softest sources. We then called RapidXMM on the position of all the sources lying in the 4XMM DR11 footprint, in order to retrieve \xmm~EPIC 0.2--12 keV flux $3\sigma$ upper limits (even in the case of non-\xmm~sources, for instance very faint \chandra~sources, or hopefully transient events). This provided us with 2.8 million flux upper limits, out of which $\sim$200 000 are constraining (i.e., lower than the minimum multi-instrument flux).

Once this archival X-ray multi-instrument catalog was built and \xmm~upper limits computed, we developed the STONKS pipeline, which takes new detections from an \xmm~observation and compares them to this catalog. The variability is defined as the pessimistic ratio between the maximum and minimum 0.1--12 keV fluxes of the source (pessimistic in the sense that the error bars are subtracted for the maximum and added for the minimum). If it is above five, a variability alert figure is built, with the long-term multi-instrument light curves, spectra (using catalog-specific band photometry), a 2'$\times$2' optical view, and a summary about the source's properties. We tested the behavior of STONKS on 483 \xmm~observations from 2021. A daily alert rate of $0.7^{+0.7}_{-0.5}$ alerts per day was obtained, with 80\% of the sources being serendipitous and 40\% not in Simbad, which is encouraging for the prospect of finding new transient events. Some of the sources of interest were analysed, including a candidate TDE, a quasar with variable spectrum, a new candidate ULX and a new candidate X-ray binary, a hard flare from an AGN and, finally, a variable AGN showing ionized absorption and a reflection component in the high state. Two confirmed TDEs that were targets of their observation were also detected, further confirming the ability of STONKS to find these variable objects. After manual screening, we estimated the false alarm rate to be below 20\%, mostly due to failures of the spectral assumption (i.e., the source is  spectrally too hard or soft). We have specifically designed the alert figure to allow us to easily visually identify this situation as well as automatically raising a warning, using the catalog-specific band photometry. The STONKS alerts should be manually screened to ensure their quality.
  
STONKS could provide the X-ray community with a new ability to detect and follow up on astrophysical transients, and possibly build synergies with future multi-wavelength transients, such as the \textit{Vera C. Rubin }Observatory for instance. This could be very useful with respect to furthering our understanding of many astrophysical transient events. The archival multi-instrument catalog  is a by-product of our method, but it can have many uses on its own. It has been made available to the community
 and will be kept up to date with ulterior data releases, including the first eROSITA sky surveys.

\begin{acknowledgements}

Softwares: \texttt{numpy} \citep{harris_array_2020}, \texttt{matplotlib} \citep{hunter_matplotlib_2007}, \texttt{astropy} 
\citep{astropy_collaboration_astropy_2013,astropy_collaboration_astropy_2018,astropy_collaboration_astropy_2022}, \texttt{astroquery} \citep{ginsburg_astroquery_2019}, \texttt{CMasher} \citep{van_der_velden_cmasher_2020}, \texttt{Xspec} \citep{arnaud_xspec_1996}, \texttt{SAS} \citep{gabriel_xmm-newton_2004}. This research has made use of \texttt{hips2fits}, a tool developed at CDS, Strasbourg, France aiming at extracting FITS images from HiPS sky maps with respect to a WCS.

The authors thank the anonymous referee for useful comments that helped improve the quality of this paper.
Some of this work was done as part of the XMM2ATHENA project. This project has received funding from the European Union's Horizon 2020 research and innovation programme under grant agreement n°101004168, the XMM2ATHENA project. EQ thanks Mickaël Coriat for his help on the web service implementation. EQ, NAW, and EK acknowledge the CNES who also supported this work. IT gratefully acknowledges the support by Deutsches Zentrum f\"ur Luft- und Raumfahrt (DLR) through grants 50\,OX\,1901 and 50\,OX\,2301.

\end{acknowledgements}

\newpage

\bibliography{STONKS.bib}

\begin{appendix} 
\section{Summary of the catalog quality filters}
\label{sec:AppSelectionCriteria}
\begin{itemize}
    \item \textbf{\textit{XMM-Newton} Pointed:}
    \begin{enumerate}
        \item "\texttt{EP\_8\_DET\_ML>8 \& SUM\_FLAG<3}", which are the detection likelihood and summary of quality flags, selected to ensure a good quality detection. The condition on the likelihood means that these detections are at a $\sim5\sigma$ level. The \texttt{SUM\_FLAG} is a summary of the various detection quality flags (see details on the \textit{XMM-Newton} catalog website\footnote{\href{http://xmmssc.irap.omp.eu/Catalogue/4XMM-DR11/col_flags.html}{http://xmmssc.irap.omp.eu/Catalogue/4XMM-DR11/col\_flags.html}}, and this value means that the detection was cleared by the manual screening;
        \item "\texttt{EP\_EXTENT==0}" (which actually means that \texttt{EP\_EXTENT} $<$6"), allowing us to exclude detections where the source is extended
    \end{enumerate}
    \item \textbf{\textit{XMM-Newton} Stacked:}
        \begin{enumerate}
            \item "\texttt{EP\_DET\_ML>8 \& STACK\_FLAG<3 \& EXTENT==0}", same criteria as for the detection catalog;
            \item "\texttt{IAUNAME\_4XMMDR11==Null}", i.e., we only keep detections which are new and not present in the detection 4XMM DR11 catalog
        \end{enumerate}
    \item \textbf{\textit{XMM-Newton} Slew:}
        \begin{enumerate}
            \item We used the Clean catalog, so "\texttt{DET\_ML>10.5}" (i.e., $\sim6\sigma$ detection) and some sources have been manually excluded;
            \item "\texttt{EXT\_B8==0 | (NULL\_EXT\_ML\_B8 \& (EXT\_B6==0. | EXT\_B7==0.))}": we only keep point-like sources, but as a detection can happen in any of the bands 8, 6 or 7, we only ask for the extent to be zero in the actual detection band.
        \end{enumerate}
    \item \textbf{\textit{Chandra}:}
        \begin{enumerate}
            \item "\texttt{likelihood\_class==TRUE}": select only good quality detections. This is based on a detection likelihood threshold that is variable and function of the observation's properties, and is computed to have at most 0.1 false detection per stacks of observations. Such good quality detections are the ones above this threshold;
            \item "\texttt{name}" does not end with "X" \& "\texttt{conf\_code<256}": removes the extended sources, and the sources that lie within the extension of another source;
            \item "\texttt{conf\_flag==FALSE}": removes detections for which the association with a single \textit{Chandra} source is ambiguous, due to off-axis PSF;
            \item Detections and upper limits are separated based on the filter "\texttt{flux\_aper\_b==0.}".
        \end{enumerate}
    \item \textbf{\textit{Swift}:}
        \begin{enumerate}
            \item We used the clean source sub-sample, so the detection quality flag is 0 or 1, the field quality flag is 0 or 1, and only datasets of quality flag 0 or 1 are used. 
            \item This catalog natively only contains sources seen as point-like for \textit{Swift};
            \item We excluded the detections where "\texttt{Rate0==0.0}"; while these might naively correspond to upper limits, the 2SXPS catalog is not supposed to contain such upper limits. These $\sim$1000 detections are thus considered as spurious, and removed;
        \end{enumerate}
    \item \textbf{ROSAT Survey:}
        \begin{enumerate}
            \item "\texttt{EXI\_ML>8 \& S\_flag==0}", to only keep good quality detections
            \item "\texttt{EXT==0.}", to only keep point-like sources. We also removed any source closer than 10' to a \textit{XMM-Newton} or \textit{Chandra} bright extended source, as some point-like sources for ROSAT are extended for better spatially resolved instruments, meaning that the source is excluded and ulterior associations are spurious;
        \end{enumerate}
    \item \textbf{ROSAT Pointed:}
            \begin{enumerate}
            \item "\texttt{Qflag>=8}", which is a summary flag allowing us to exclude any extended source, located-within-extension source, or any type of spurious detection;
            \item We also removed any source closer than 10' to a \textit{XMM-Newton} or \textit{Chandra} bright extended source;
        \end{enumerate}
    \item \textbf{eROSITA:}
        \begin{enumerate}
            \item "\texttt{DET\_LIKE>8}" (i.e., $\sim5\sigma$ detection), to keep only good quality detections;
            \item "\texttt{EXT==0.}", exclude extended sources
        \end{enumerate}
\end{itemize}

\clearpage 

\onecolumn 
\section{Energy conversion factors}
~

\begin{table*}[h]
\centering
\begin{tabular}{c||c|c||c|c||c|c}
Catalog & Total Band & Total fraction & Soft band & Soft threshold &  Hard band & Hard threshold\\ \hline \hline
XMM-DR11, DR11s, SL2 & 0.2--12 keV& 0.999 & 0.2--2 keV & <-0.42 & 2--12 keV & >0.88\\
2SXPS & 0.3--10 keV& 0.9 & 0.3--2 keV& <-0.4 & 2--10 keV & >0.84\\
CSC2 & 0.5--7 keV&0.69  &0.5--2 keV & <-0.33 & 2--7 keV & >0.774\\
eFEDS & 0.2--4.5 keV& 0.60 & 0.2--2 keV & <-0.62 & 2--4.5 keV& >0.45\\
RASS, WGACAT & 0.1--2.4 keV& 0.35 & 0.2--2.4 keV & N/A & N/A & N/A\\
\end{tabular}
\caption{The various total, soft and hard energy bands of the catalogs considered in this work. For the total band, we indicate the fraction of reference total flux (0.1--12 keV for a spectrum with $\Gamma=1.7$ and $N_{\rm H}=3\times10^{20}$~cm$^{-2}$) this band contains. This allows us to calibrate the various catalogs, assuming this underlying spectral shape. For the soft and hard bands, we show the threshold in hardness ratio above (resp. below) which a detection is in the 10 \% hardest (resp. softest) of its catalogs, which could lead to errors of factor of $\sim2$ in the flux calibration and, thus, in the variability computation.}
\label{tab:EnergyConvFactor}
\end{table*}

\section{STONKS alert}
\subsection{Statistics}
\begin{figure*}[h]
    \centering
    \includegraphics[width=\textwidth]{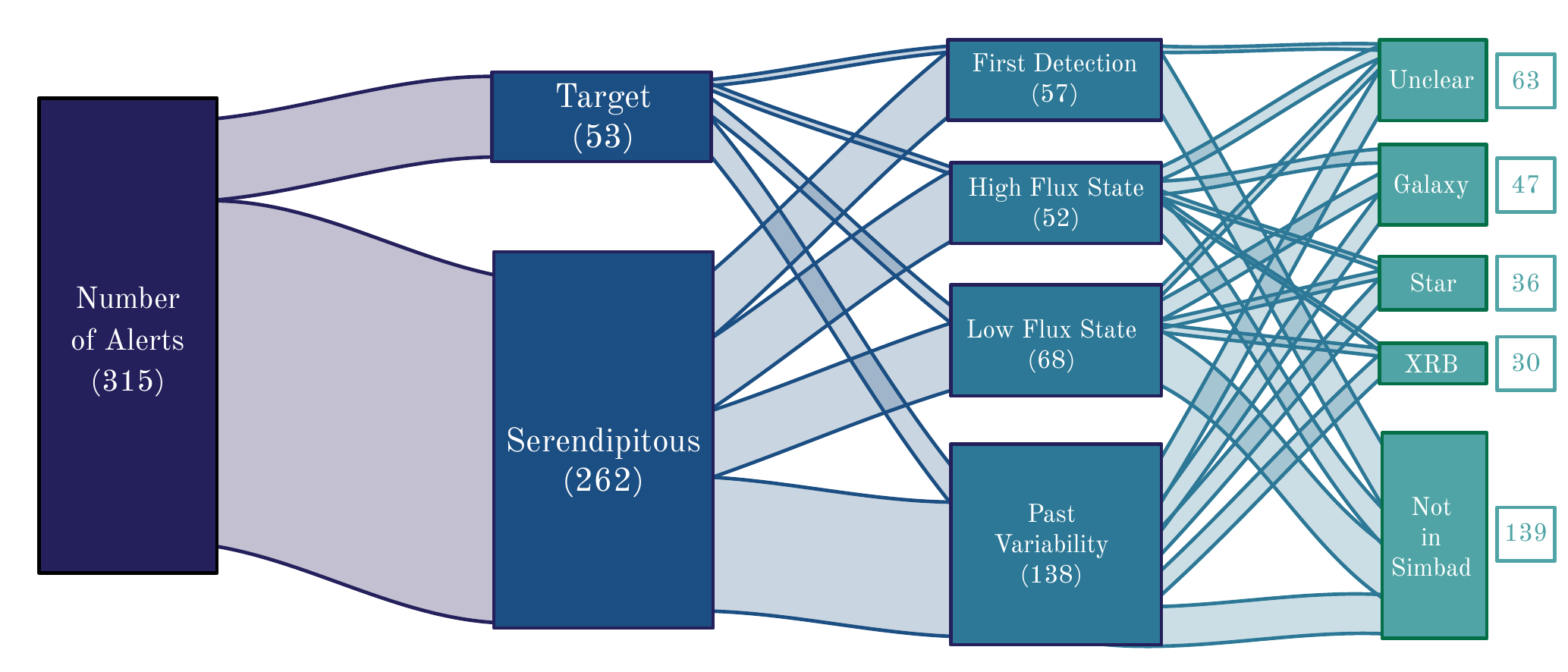}
    \caption{Statistics of the test run of STONKS on a part of the 2021 \xmm~archive. The height of the boxes and branches are proportional to the number of alerts -- we have chosen to not display the exact numbers for the sake of readability. The main takeaways are the high fraction of serendipitous alerts and the high fraction of sources that are either 1) not in Simbad or  2) in Simbad, but with no details on the nature of the object. This shows the potential of STONKS to uncover new hiddent transients.}
    \label{fig:StatAlerts}
\end{figure*}

\clearpage 

\onecolumn
\subsection{Alerts from sources of interest from the 2021 STONKS test run} 
\label{sec:AppTesting}
\begin{figure*}[!htb]
    \centering
    \includegraphics[width=\textwidth]{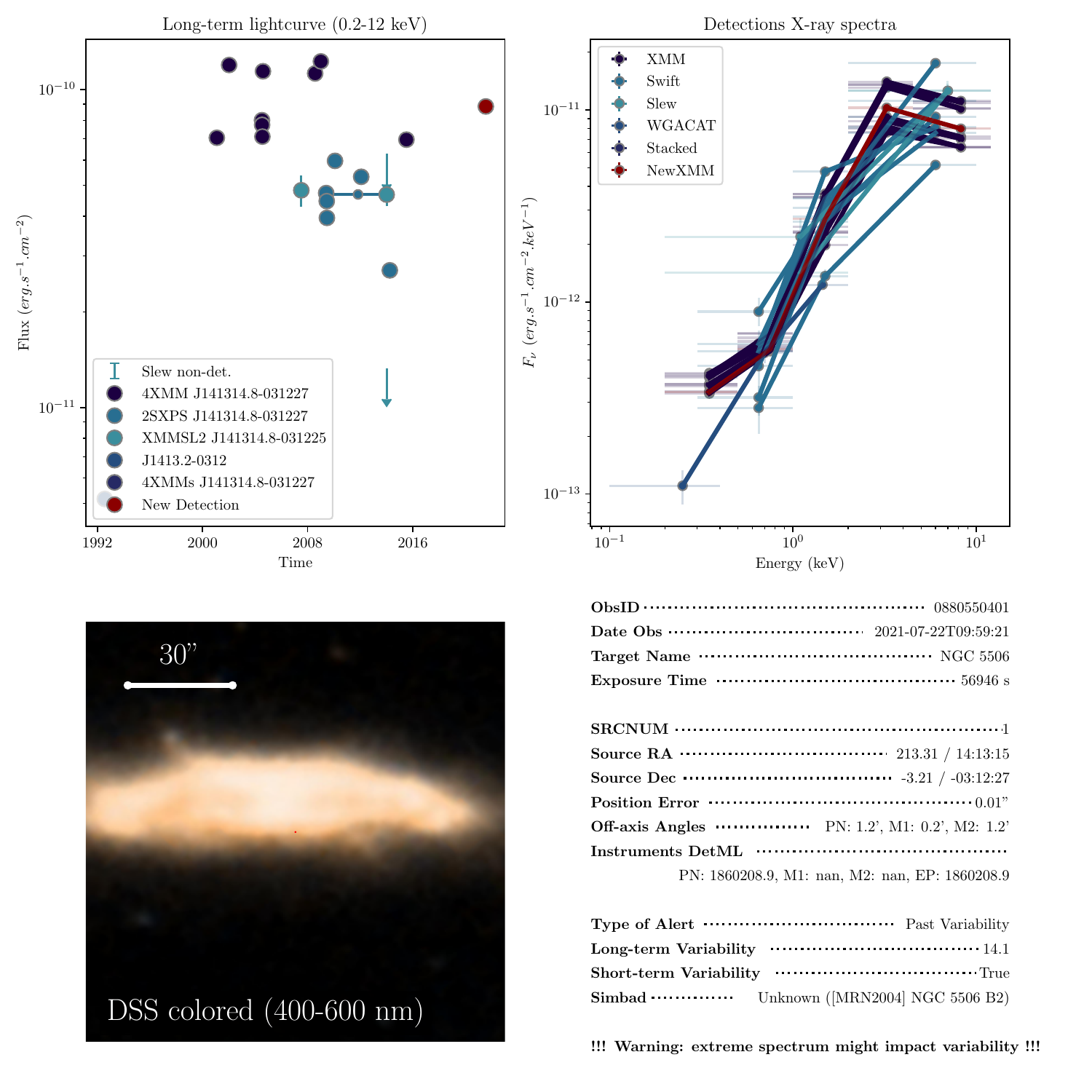}
    \caption{Example of spurious variability due to the hardness of the source (here, due to the amount of absorption in the host galaxy). The tiny red dot in the middle of the DSS image (bottom left) is the $1\sigma$ positional error circle of the X-ray source.}
    \label{fig:SpectralErrorAlert}
\end{figure*}
\begin{figure*}
    \centering
    \includegraphics[width=\textwidth]{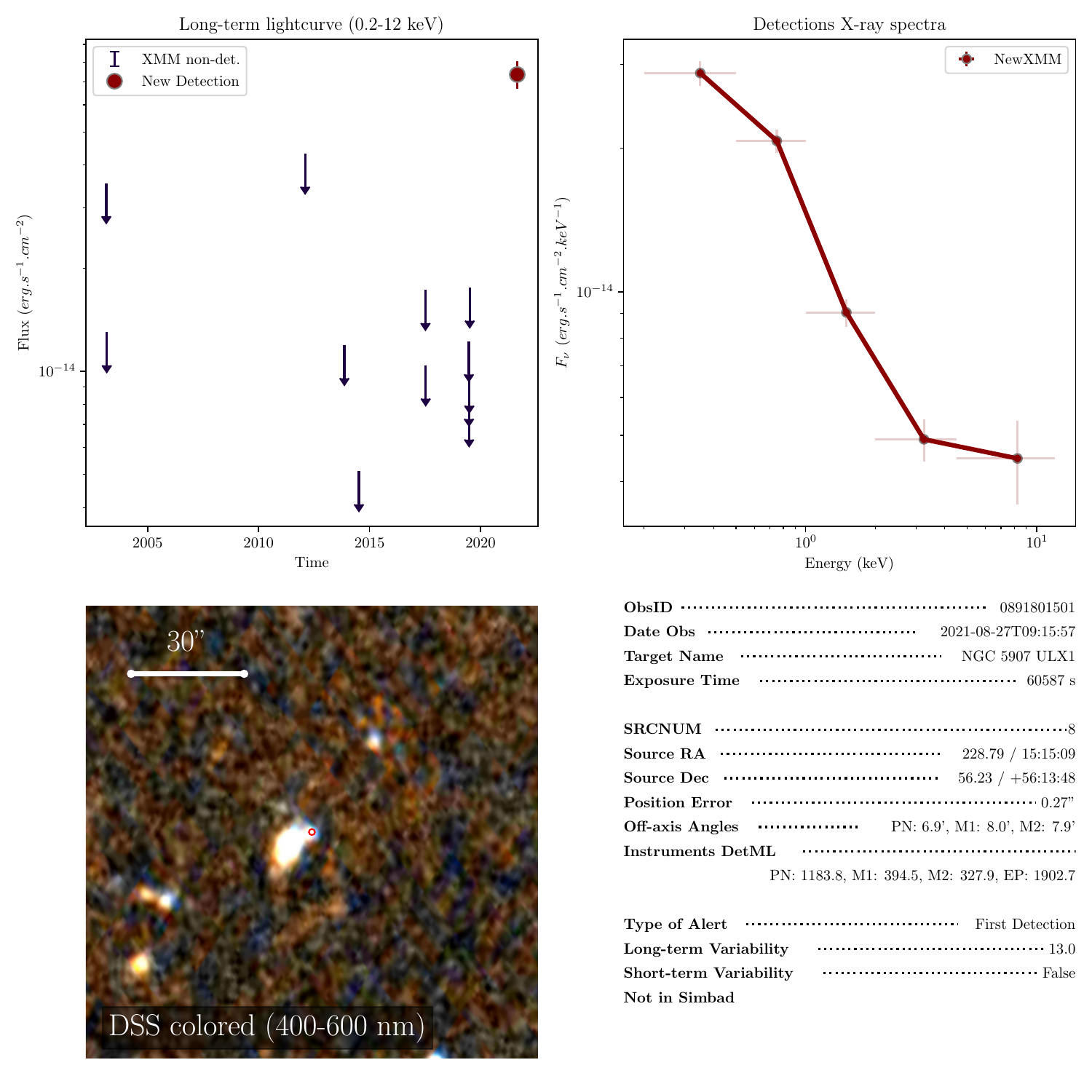}
    \caption{Example of an alert sent out by STONKS: a possible TDE candidate or a flaring AGN. }
    \label{fig:Tde_Alert}
\end{figure*}
\begin{figure*}
    \centering
    \includegraphics[width=\textwidth]{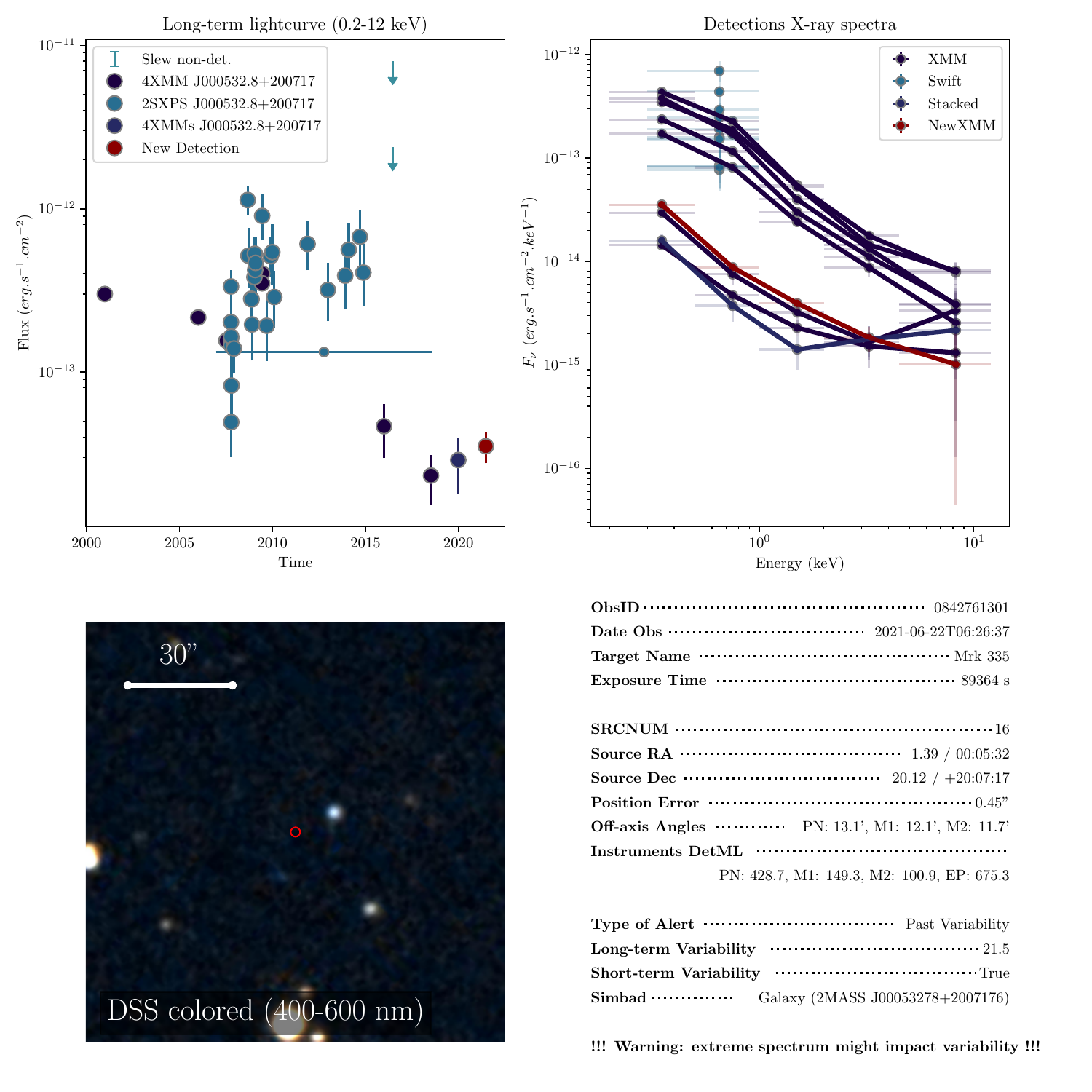}
    \caption{Example of an alert sent out by STONKS: a quasar with variable photon-index. }
    \label{fig:QuasarSpecVar_alert}
\end{figure*}
\begin{figure*}
    \centering
    \includegraphics[width=\textwidth]{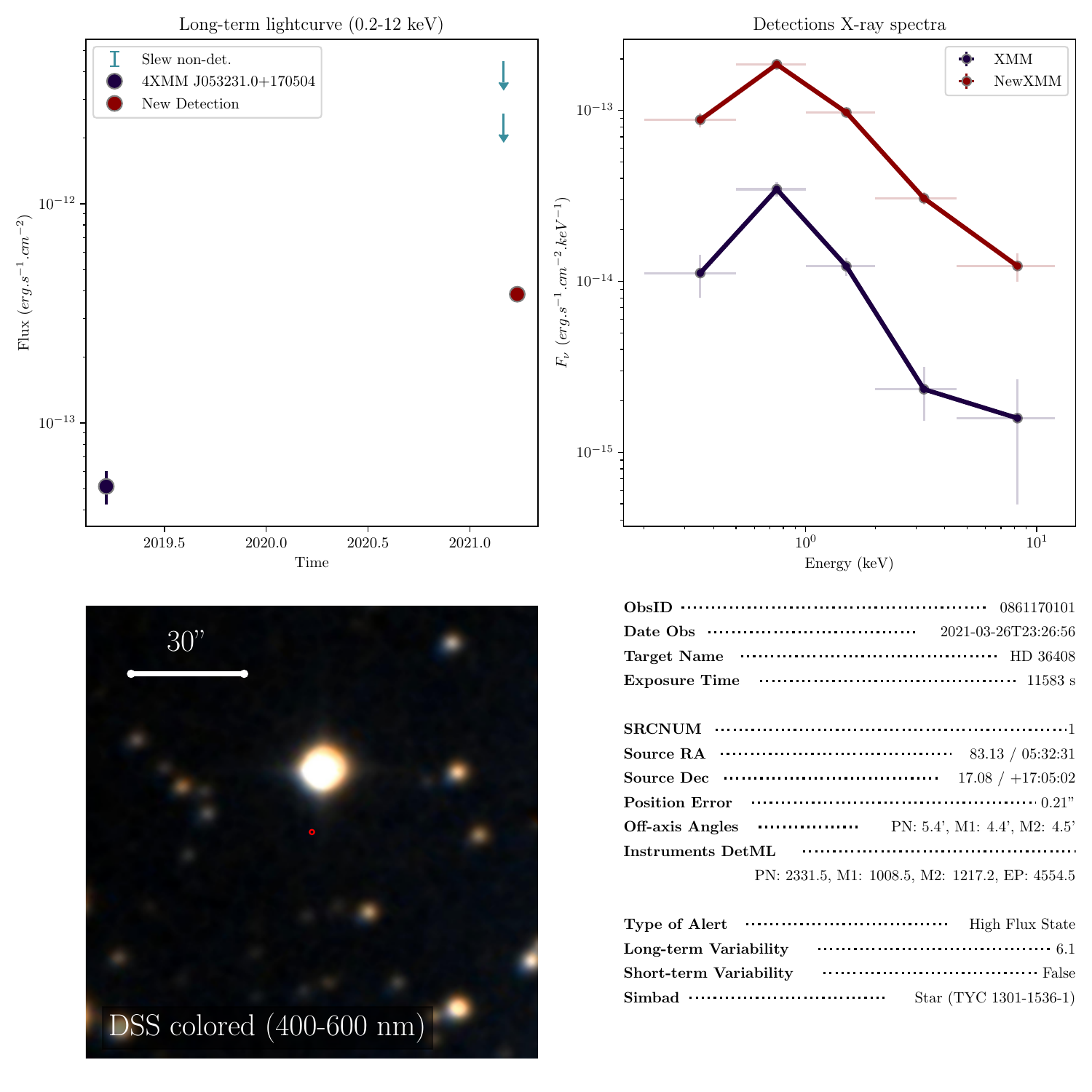}
    \caption{Example of an alert sent out by STONKS: a stellar flare.}
    \label{fig:StellarFlare_alert}
\end{figure*}
\begin{figure*}
    \centering
    \includegraphics[width=\textwidth]{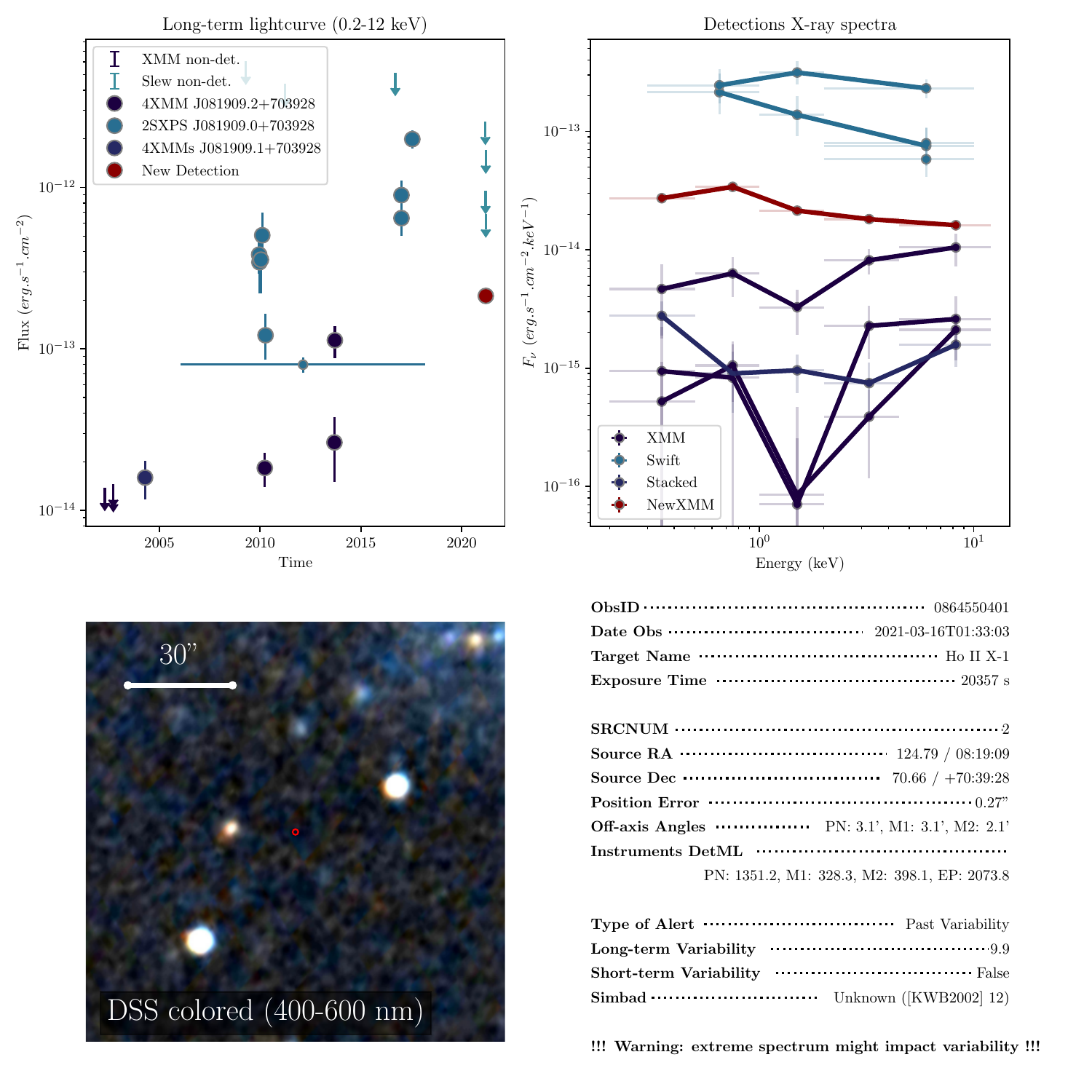}
    \caption{Example of an alert sent out by STONKS: a possibly mis-classified ULX candidate. }
    \label{fig:MisclassifiedULX}
\end{figure*}
\begin{figure*}
    \centering
    \includegraphics[width=\textwidth]{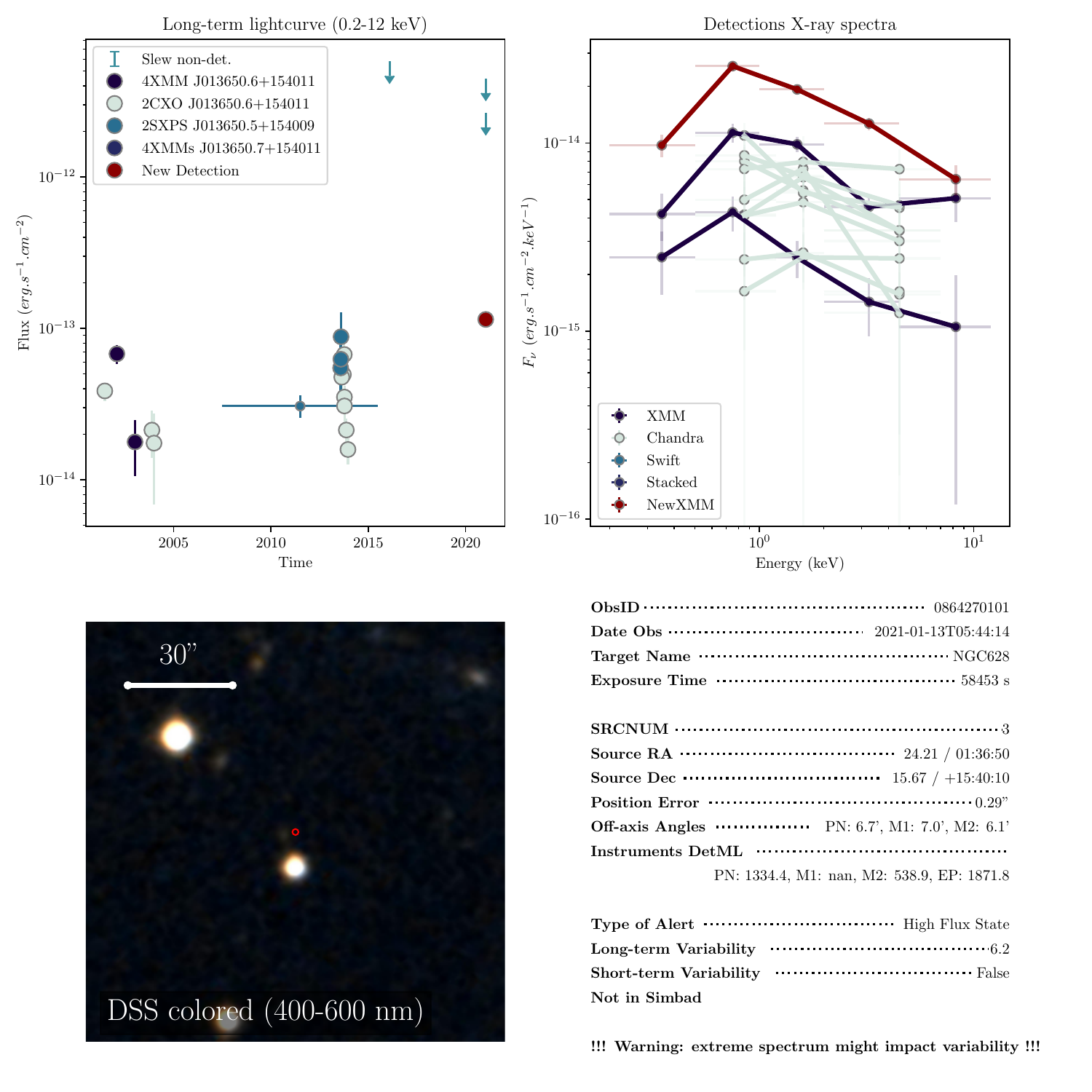}
    \caption{Example of an alert sent out by STONKS: a new candidate XRB. }
    \label{fig:NewXRB}
\end{figure*}

\begin{figure*}
    \centering
    \includegraphics[width=\textwidth]{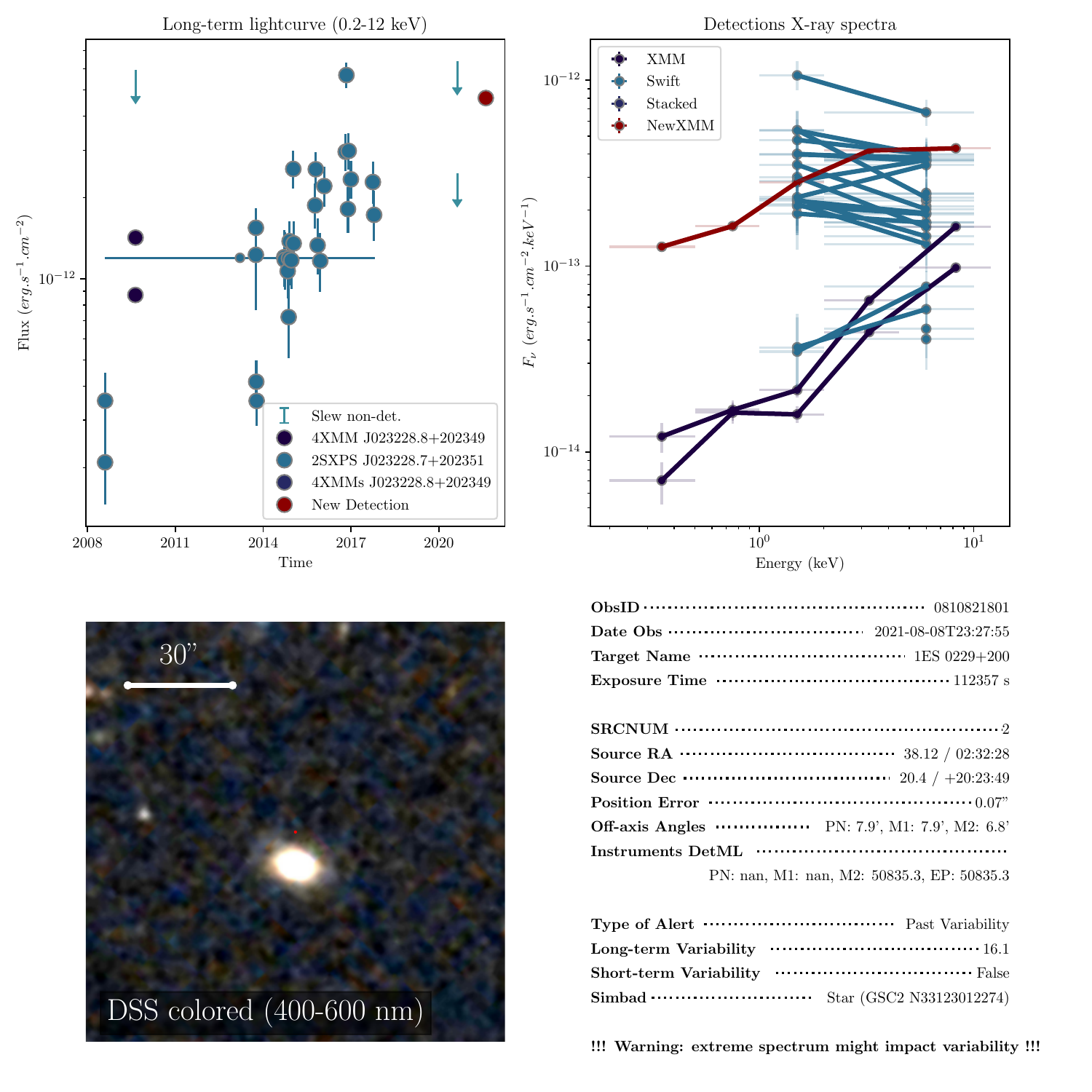}
    \caption{Example of an alert sent out by STONKS: a short-term variable AGN with ionized absorption.}
    \label{fig:AbsorbedAGN}
\end{figure*}

\FloatBarrier
\twocolumn
\subsection{Spectra from sources of interest from the 2021 STONKS test run} 
\begin{figure}[h]
    \centering
    \includegraphics[width=\columnwidth]{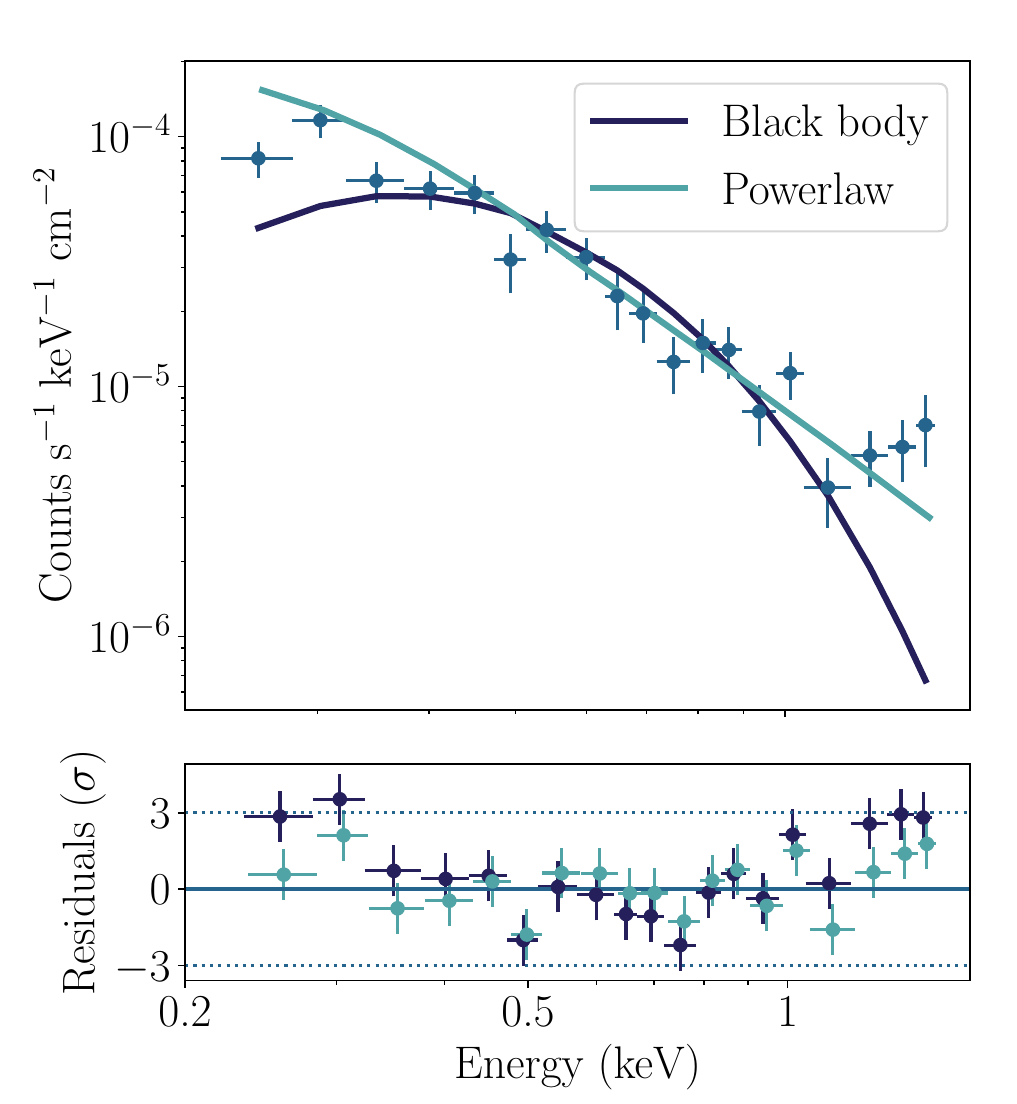}
    \caption{\xmm~EPIC pn spectrum of the TDE-like flare of 4XMM\,J151509.5+561347, with two models (absorbed powerlaw or absorbed black body).}
    \label{fig:TDEspectrum}
\end{figure}
\begin{figure}[h]
    \centering
    \includegraphics[width=\columnwidth]{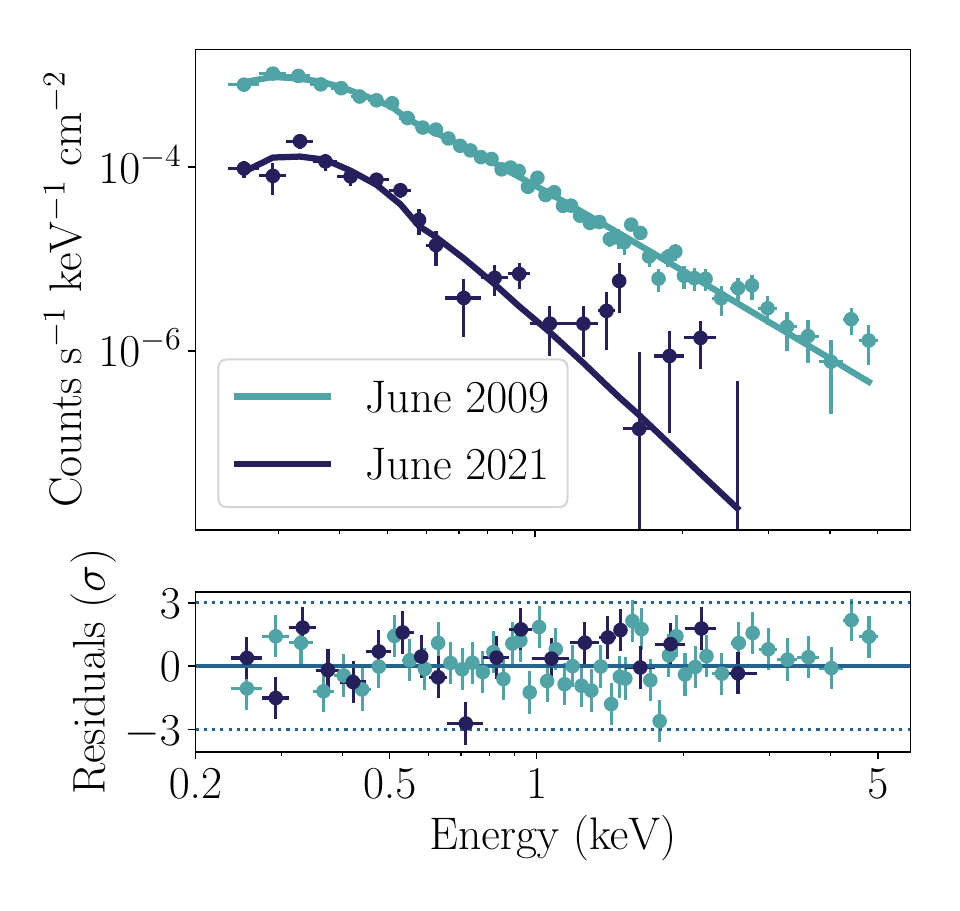}
    \caption{\xmm~EPIC pn spectrum of the variable quasar 4XMM\,J000532.8+200717, fitted with an absorbed powerlaw model.}
    \label{fig:Quasar_SpecVar_Spectra}
\end{figure}
\begin{figure}[h]
    \centering
    \includegraphics[width=\columnwidth]{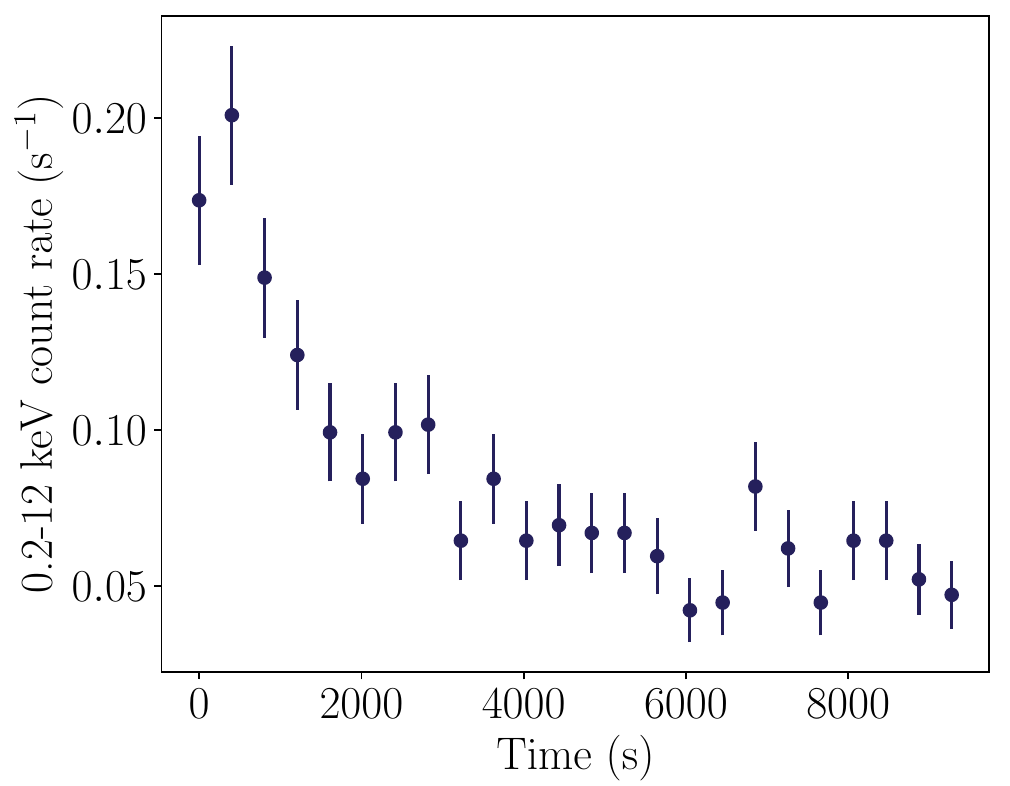}
    \caption{\xmm~EPIC pn 0.2-12 keV lightcurve of the flaring star 4XMM\,J053231.0+170504.}
    \label{fig:StellarFlareLightcurve}
\end{figure}
\begin{figure}[h]
    \centering
    \includegraphics[width=\columnwidth]{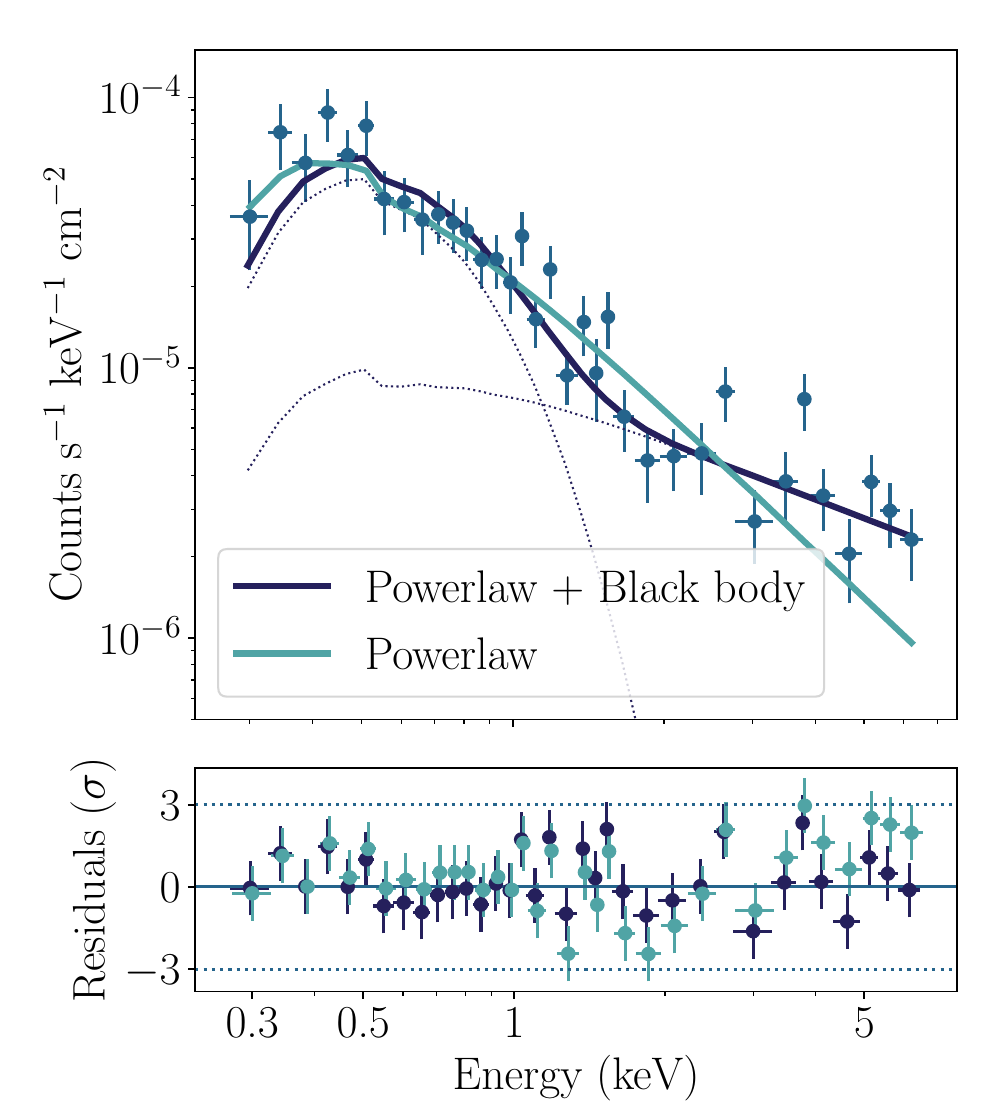}
    \caption{\xmm~EPIC pn spectrum of 4XMM J081909.2+703928}
    \label{fig:MisclassifiedULXspectrum}
\end{figure}
\begin{figure}[h]
    \centering
    \includegraphics[width=\columnwidth]{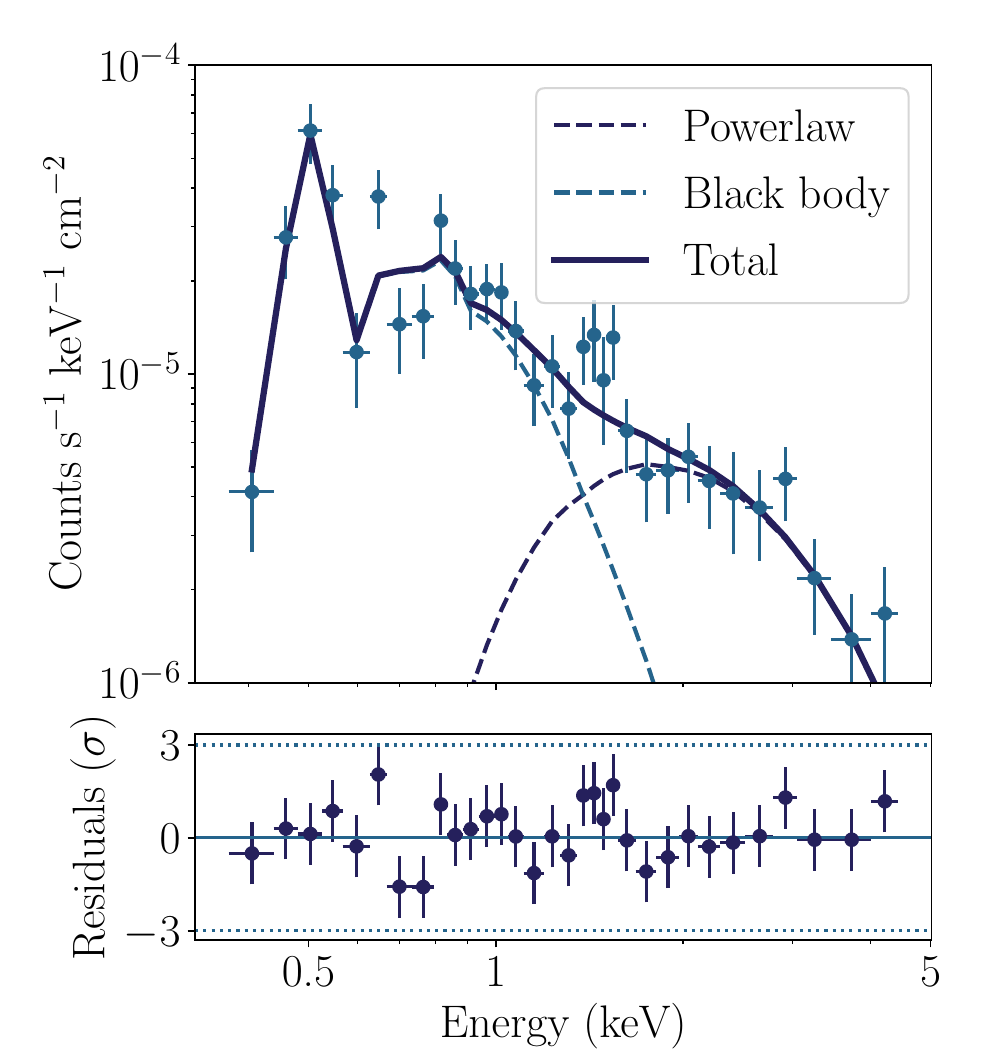}
    \caption{\xmm~EPIC pn spectrum of 4XMM J013650.6+154011}
    \label{fig:XRBspectrum}
\end{figure}


\begin{figure}[h]
    \centering
    \includegraphics[width=\columnwidth]{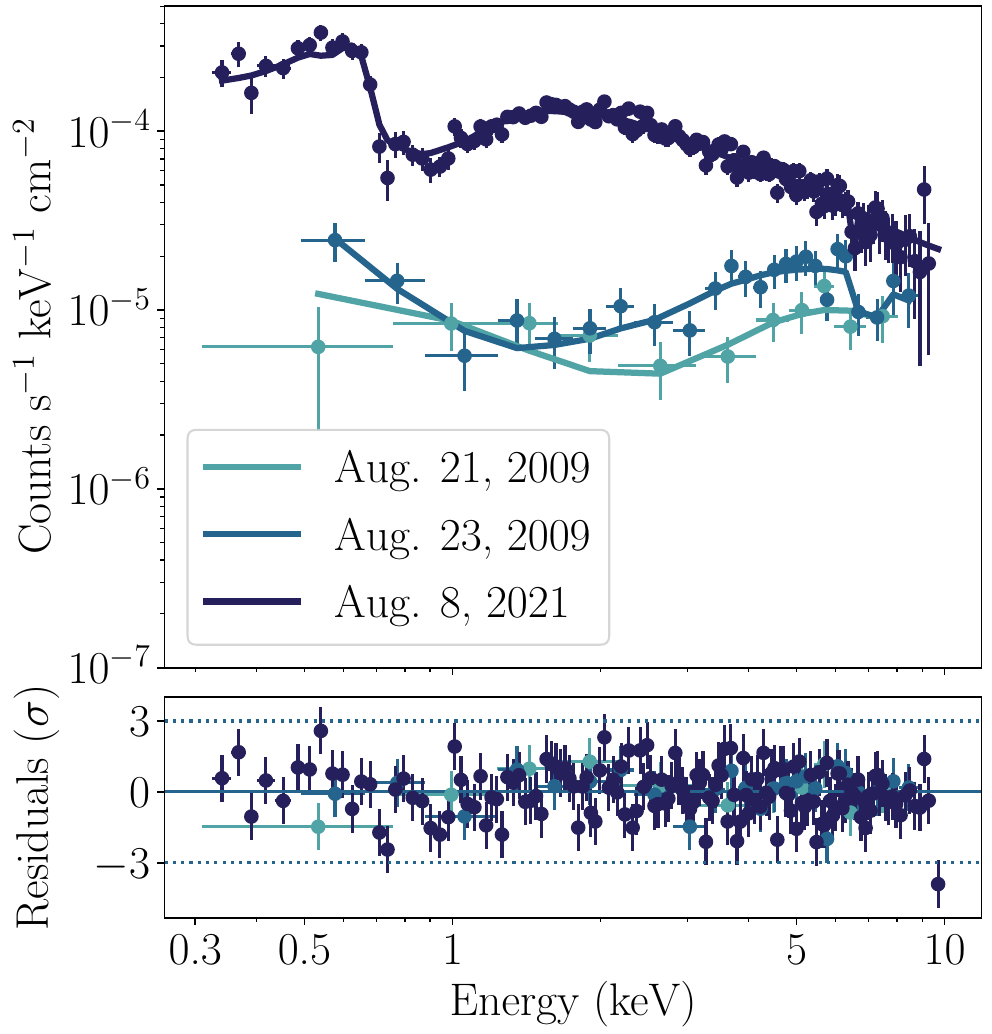}
    \caption{\xmm~EPIC pn spectra of 4XMM J023228.8+202349 from three different observations. The spectra show a variable powerlaw emission, with ionized absorption and reflection.}
    \label{fig:HardIonizedAbs_spectrum}
\end{figure}

\begin{figure}[h]
    \centering
    \includegraphics[width=\columnwidth]{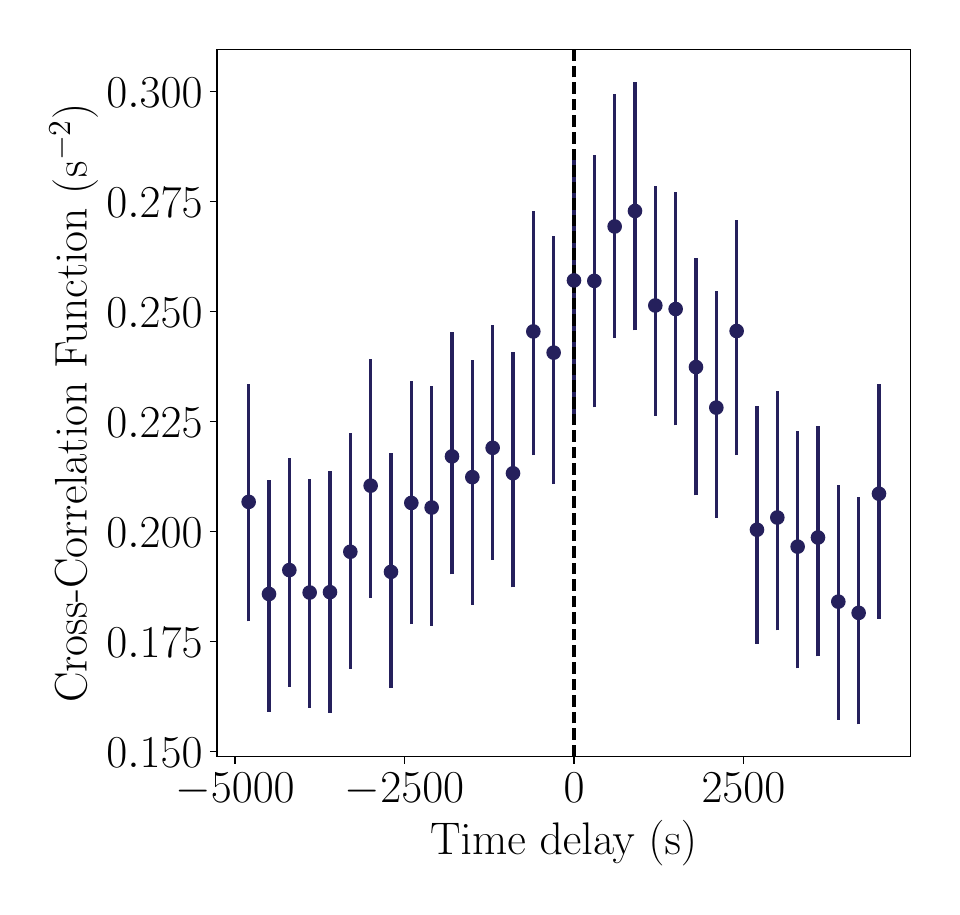}
    \caption{Cross-correlation function of 4XMM J023228.8+202349 (ObsID 0810821801), showing the lag between the soft (0.3--2 keV) and hard (2--7 keV) bands. The cross-correlation function corresponds to CCF$(\tau) = \left< \left( F_{\rm soft}(t+\tau)-\bar{F}_{\rm soft}\right)\times \left( F_{\rm hard}(t)-\bar{F}_{\rm hard}\right) \right>$ \citep[e.g.,][]{white_comments_1994}. The lightcurves were binned at 300s.}
    \label{fig:AGN_lag}
\end{figure}
\newpage 

\newpage

\FloatBarrier
\onecolumn
\section{Flux comparison between matched catalogs}
\begin{figure*}[h]
    \centering
    \includegraphics[angle=270, width=0.8\textwidth]{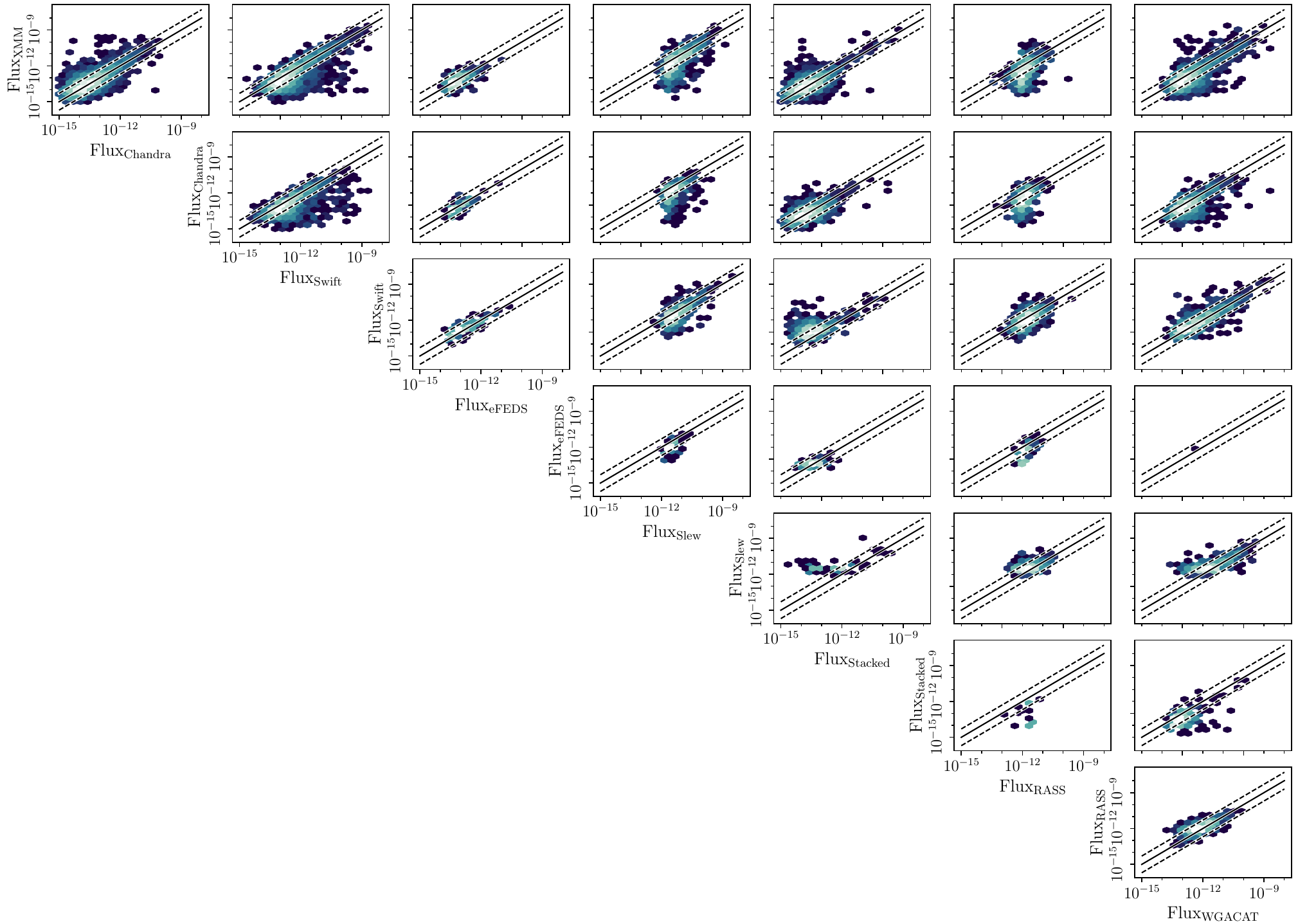}
    \caption{Two-by-two flux comparisons of the various catalogs within our archival cross-matched catalog. Each flux value is the average over all the catalog-specific detections, to avoid the bias towards variable sources being more observed. All fluxes are given in erg s$^{-1}$ cm$^{-2}$, after extrapolation to the 0.1--12 keV band as explained in Sect. \ref{subsec:FluxCal}.}
    \label{fig:FluxvsFlux}
\end{figure*}


\end{appendix}

\end{document}